\documentclass{aa} 

\newcommand{\BCGone}{3C~244.1~}
\newcommand{\BCGtwo}{SDSS~J161112.65+550823.5~}

\usepackage{lscape}

\usepackage{changepage}
\usepackage{graphicx}
\usepackage{subfig}
\usepackage{txfonts}
\usepackage{natbib}
\usepackage{comment}
\DeclareGraphicsExtensions{.pdf,.png,.jpg}
\bibpunct{(}{)}{;}{a}{}{,} 
\begin{document}

   \title{Molecular gas in distant brightest cluster galaxies}

   \author{G. Castignani
          \inst{1}\fnmsep\thanks{e-mail: gianluca.castignani@epfl.ch}
          \and
          F. Combes\inst{2,3}
          \and
          P. Salom\'e\inst{2}
          \and
          J. Freundlich\inst{4}
          }

   \institute{Laboratoire d'astrophysique, \'{E}cole Polytechnique F\'{e}d\'{e}rale de Lausanne (EPFL), Observatoire de Sauverny, 1290 Versoix, Switzerland
              \and
         Sorbonne Universit\'{e}, Observatoire de Paris, Universit\'{e} PSL, CNRS, LERMA, F-75014, Paris, France
               \and
             Coll\`{e}ge de France, 11 Place Marcelin Berthelot, 75231 Paris, France
             \and 
            Centre for Astrophysics and Planetary Science, Racah Institute of Physics, The Hebrew University, Jerusalem 91904, Israel
              }
                            \date{Received June 21, 2019; Accepted 23 December 2019}

  \abstract
   {The mechanisms governing the stellar mass assembly and star formation history of brightest cluster galaxies (BCGs) are still being debated. By means of new and archival molecular gas observations we investigate the  role of dense megaparsec-scale environments in regulating the fueling of star formation in distant BCGs, through cosmic time.
   We  observed in CO with the IRAM 30m telescope  two star-forming BCGs  belonging to SpARCS clusters, namely, \BCGone ($z=0.4$) and \BCGtwo ($z=0.9$), and compared their molecular gas and star formation properties with those of a compilation of $\sim100$ distant cluster galaxies from the literature, including nine additional distant BCGs at ${z\sim0.4-3.5}$.
   We  set robust upper limits of $M_{{\rm H}_2}<1.0\times10^{10}~M_\odot$ and $<2.8\times10^{10}~M_\odot$ to their molecular gas content, respectively, and  to the ratio of  molecular gas to stellar mass  $M({\rm H}_2)/M_\star\lesssim0.2$ and depletion time $\tau_{\rm dep}\lesssim40$~Myr of the two targeted BCGs. They are thus among the distant cluster galaxies with the lowest gas fractions and shortest depletion times. The majority ($64\%\pm15\%$ and $73\%\pm18\%$) of the 11 BCGs with observations in CO have lower $M({\rm H}_2)/M_\star$ values and $\tau_{\rm dep}$, respectively, than those estimated for main sequence galaxies. Statistical analysis also tentatively suggests that the values of $M({\rm H}_2)/M_\star$ and $\tau_{\rm dep}$ for the 11 BCGs deviates, with a significance of $\sim2\sigma$, from those of the comparison sample of cluster galaxies.  A morphological analysis  for a subsample of seven BCGs with archival {\it HST} observations reveals that $71\%\pm17\%$ of the BCGs are compact or show star-forming components or substructures.
   Our results suggest a scenario where distant star-forming BCGs assemble a significant fraction $\sim16\%$ of their stellar mass on the relatively short timescale $\sim\tau_{\rm dep}$, while environmental mechanisms might prevent the replenishment of gas feeding the star formation.  We speculate that compact components also favor the rapid exhaustion of molecular gas and ultimately help to quench the BCGs.
Distant star-forming BCGs  are excellent targets for ALMA and for   next-generation telescopes
such as the {\it James Webb Space Telescope}.}


   \maketitle
%

\section{Introduction}\label{sec:introduction}

Brightest cluster galaxies (BCGs) are excellent laboratories for studying the effect of dense galaxy cluster environments on galaxy evolution. 
In the local Universe they are commonly associated with passively evolving massive ellipticals of cD type, which often host radio galaxies \citep{Zirbel1996}, and are located at the center of the cluster cores \citep{Lauer2014}, where tight correlations involving galaxy properties such as morphology \citep{Dressler1980}, color \citep{Kodama2001}, stellar mass 
\citep{Ostriker_Tremaine1975}, star formation \citep{Peng2010}, and gas content \citep{Gunn_Gott1972} are observed. 
Because of their exceptional masses and luminosities and {because they are  located in the {crowded} cores of clusters}, BCGs are believed to evolve via phenomena such as dynamical friction \citep{White1976}, galactic cannibalism \citep{Hausman_Ostriker1978}, interactions with the intracluster medium \citep{Stott2012}, and cooling flows \citep{Salome2006}.

{ How the growth and star formation history of BCGs are regulated is still a matter of debate.} Recent work suggests that they have doubled their stellar mass since $z\sim1$ \citep{Lidman2012}, { consistent} with a global picture where BCGs evolve via dry accretion of satellite galaxies 
\citep{Collins2009,Stott2011}. More recent studies, however,  have  found potentially conflicting results to this somewhat simplistic hypothesis. 
Possible evidence for high star formation rates and large reservoirs of molecular gas in BCGs is found out to $z\sim1$ and beyond \citep{McDonald2016,Webb2015,Webb2015b,Bonaventura2017}, thus { favoring} a scenario where star formation is fed by rapid gas deposition at high-$z$ and slow cooling flows at low-$z$
\citep{Ocvirk2008,Dekel2009a,Dekel2009b}.

Several studies have also found a high concentration of potentially in-falling, star-forming galaxies in the outskirts
of nearby clusters \citep[e.g.,][]{Bai2009,Chung2010}, and a strong 
{ increase in  the fraction of star-forming galaxies}
in cluster cores out to $z\sim1$ and beyond \citep{Smith2010,Tran2010,Tadaki2012,Webb2013,Brodwin2013,Zeimann2013,Santos2015,Alberts2016, Wang2016}. Such findings seem to {support} the late assembly of cluster core members via  the infall of gas-rich systems and via strong environmental quenching mechanisms
\citep[e.g., strangulation, ram pressure stripping, and galaxy harassment][]{Larson1980,Moore1999}.

On the other hand, other studies suggest the presence of a significant population of quiescent galaxies even in $z\gtrsim1.5$ cluster cores \citep{Tanaka2013,Koyama2014, Newman2014,Cooke2016,Strazzullo2016,Strazzullo2019}.
In particular, \citet{Strazzullo2019} have recently studied the cluster galaxy population, including the BCGs, of a sample of five clusters at $z=1.4-1.7$ from the { South-Pole Telescope Sunyaev Zel’dovich effect (SPT-SZ)} survey \citep{Bleem2015} and suggested that the star formation is suppressed in the cores of the massive clusters in their sample already earlier than $z\sim1.5$.

{Since molecular gas is an excellent tracer { of ongoing or future star formation} \citep{Bigiel2008,Schruba2011,Leroy2013}, observations of CO in both local and distant BCGs are a powerful tool that can be used to better understand  the mass assembly and gas fueling of these sources. }
Some studies of BCGs in the local universe have, remarkably, found molecular gas reservoirs \citep{Edge2001,Salome_Combes2003,Hamer2012,McNamara2014, Russell2014,Tremblay2016,Fogarty2019} and filaments of cold gas \citep{Olivares2019,Russell2019}.  

{ In the
literature, however,  in the more distant Universe (i.e., $z>0.4$) there are  only a handful of BCGs observed in CO \citep[e.g.,][]{Webb2017,Emonts2013}, mostly at $z>1$, where the average ratio of molecular gas to stellar mass  in galaxies, at least in the field, is expected to increase significantly (by a factor of $\gtrsim4$) with respect to local galaxies \citep{Carilli_Walter2013}.} { To probe the star formation fueling of cluster core galaxies we  { recently} performed a wide search of molecular gas in distant BCGs \citep{Castignani2019} and cluster galaxies \citep{Castignani2018} over a wide range of redshifts: $z\sim0.4-2.6$. As part of this search, in this work we report CO observations of { two other} distant star-forming BCGs. The two BCGs span a broad range in cosmic time (3~Gyr), being located at $z=0.4$ and $z=0.9$, within which the overall molecular gas content of galaxies \citep{Carilli_Walter2013} and their star formation activity \citep{Madau_Dickinson2014} are expected to increase by a factor of $\sim2$. With this work we thus complement our recent studies to reveal the fueling of gas feeding the star formation} in the high-$z$ { counterparts}  of present day star-forming ($>40~M_\odot$/yr) BCGs, such as the famous Perseus~A and Cygnus~A galaxies \citep{FraserMcKelvie2014}.


In this work we  refer to proto-BCGs, BCGs, and BCG candidates with no distinction, keeping in mind that the secure identification of distant BCGs is difficult. This  is ultimately due to the difficulty in confirming and characterizing high-$z$ cluster members and in particular the BCGs, which  reside in { crowded} regions of the cluster cores and often have a complex (multi-component) morphology.

The paper is  structured as follows: in Sect.~\ref{sec:sample} we describe the two targets; in Sect.~\ref{sec:observations_and_data_reduction} we describe the observations and data reduction; in Sect.~\ref{sec:results} we present the results; in Sect.~\ref{sec:summary_conclusions} we summarize the results and draw our conclusions. { In Appendix~\ref{app:appendix} we list the properties of distant cluster galaxies observed in CO.}
Throughout this work we adopt a flat $\Lambda \rm CDM$ cosmology with matter density $\Omega_{\rm m} = 0.30$, dark energy density $\Omega_{\Lambda} = 0.70$, and Hubble constant $h=H_0/(100\, \rm km\,s^{-1}\,Mpc^{-1}) = 0.70$ \citep[but see][]{PlanckCollaborationVI2018,Riess2019}.

\begin{table*}[]\centering
\begin{adjustwidth}{-0.cm}{}
\begin{center}
\begin{tabular}{ccccccccc}
\hline\hline
Galaxy ID & R.A. & Dec. & $z_{spec}$ & $L_{\rm IR}$ & $M_\star$ & SFR & sSFR & sSFR$_{\rm MS}$ \\
   & (hh:mm:ss.s) & (dd:mm:ss.s) &  & ($10^{11}~L_\odot$) & ($10^{11}~M_\odot$) & ($M_\odot$/yr) & (Gyr$^{-1}$) & (Gyr$^{-1}$)\\ 
  (1) & (2) & (3) & (4) & (5) & (6) & (7) & (8) & (9) \\
 \hline
 3C~244.1 & 10:33:34.0 & +58:14:35.5   & 0.430 & $16.1\pm1.2$   & 1.0 & $281\pm218$ & $2.8^{+2.6}_{-2.5}$ & 0.12 \\ 
 SDSS~J161112.65+550823.5 & 16:11:12.7 & +55:08:23.6 & 0.907 & $43.7\pm15.8$  & 1.8 & $766\pm275$ & $4.3^{+2.6}_{-2.9}$ & 0.30 \\ 
 \hline
\end{tabular}
 \end{center}
\caption{Properties of our targets, from \citet{Webb2015}. (1) galaxy name; (2-3) J2000 equatorial coordinates; (4) spectroscopic redshift; (5) total infrared luminosity inferred from 24~$\mu$m observer frame Spitzer MIPS fluxes and using the \citet{Chary_Elbaz2001} model;
(6) stellar mass estimated using 3.6~$\mu$m observer frame Spitzer IRAC fluxes, \citet{Bruzual_Charlot2003} stellar population
modeling, and typical rest-frame K-band mass-to-light ratio of red galaxies \citep{Bell2003}; (7) SFR estimated from $L_{\rm IR}$ using the \citet{Kennicutt1998} relation; (8) specific SFR determined as ${\rm sSFR}={\rm SFR}/M_\star$; (9) sSFR for main sequence field galaxies with redshift and stellar mass of our targets estimated using the relation found by \citet{Speagle2014}.}
\label{tab:BCG_properties}
\end{adjustwidth}
\end{table*}

\section{Two brightest cluster galaxies}\label{sec:sample}
{ We consider the \citet{Webb2015} BCG catalog which comprises observed infrared (IR) properties of a large sample of 535 BCGs within the redshift range $0.2<z<1.8$. These BCGs belong to clusters drawn from the 
Spitzer Adaptation of the Red-Sequence  Cluster  Survey (SpARCS), which is an optical--near-infrared galaxy-selected cluster survey, whose goal is to discover distant clusters out  to $z\sim2$ \citep{Muzzin2009,Muzzin2012,Wilson2009,Demarco2010}.}

{ The authors provide both SFRs and stellar masses of the BCGs. The former were estimated from 24~$\mu$m {\it Spitzer}-MIPS fluxes using the models by \citet{Chary_Elbaz2001} and the \citet{Kennicutt1998} relation. The stellar masses were instead estimated from the observed 3.6~$\mu$m flux, converted into the rest-frame K-band luminosity, taking the K-correction into account.}

As we want to target actively star-forming BCGs in order to investigate their molecular gas content and explore their evolution and interaction with their Mpc-scale environment, we consider spectroscopically confirmed sources from the \citet{Webb2015} BCG catalog. {  This selection leaves us with 16 spectroscopically confirmed sources out of the 535 BCGs of the catalog.} { We also consider the subsample of five BCGs, at $0.4<z<1.1$, with the strongest star formation activity} (i.e., star formation rate ${\rm SFR}>250~M_\odot$/yr). { A sixth BCG at $z = 1.7$ satisfies the selection requirements, but is not included because it has been already observed in CO(2$\rightarrow$1) by \citet{Webb2017}, as discussed later in this work.}

{ Of these five BCGs, we limited ourselves to two targets,} namely \BCGone ($z = 0.4$) and \BCGtwo ($z=0.9$), with SFR$=281~M_\odot$/yr and 766~$M_\odot$/yr, respectively, { for which we predicted to detect CO on the basis of standard relations by \citet{Tacconi2018}, valid for main sequence { (MS)} field galaxies.}
The two BCGs belong 
to $M_{200}\gtrsim1\times10^{14}~M_\odot$ clusters and are located in the Lockman and ELAIS-N1 fields, respectively \citep{Webb2015}. In particular, \BCGone is a powerful Type~II Fanaroff-Riley (FR) radio galaxy \citep{Fanaroff_Riley1974} hosted by a cluster with an estimated richness of 15 galaxies within the cluster core 
\citep{Hill_Lilly1991}.
{ Active galactic nucleus (AGN) contamination to the SFR estimates cannot be excluded, since the 24~$\mu$m fluxes are associated with rest-frame emission at shorter wavelengths of 17 and 13$\mu$m, respectively (see Sect.~\ref{sec:AGNcontamination_SFR} for further discussion).}

\begin{figure*}[]\centering
\captionsetup[subfigure]{labelformat=empty}
\subfloat[\BCGone]{\hspace{0.cm}\includegraphics[trim={0cm 0cm 0cm 
0cm},clip,width=0.3\textwidth,clip=true]{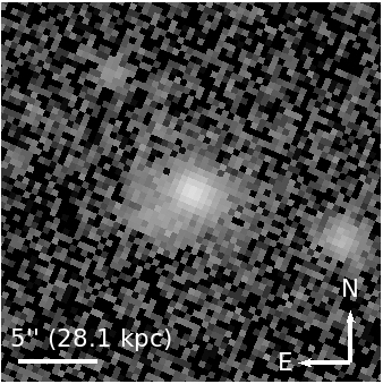}}
\subfloat[\hspace{0.7cm}\BCGtwo]{\hspace{0.7cm}\includegraphics[trim={0cm 0cm 0cm 
0cm},clip,width=0.3\textwidth,clip=true]{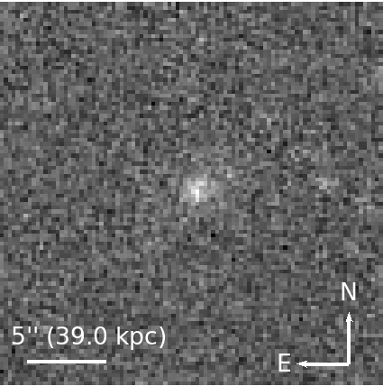}}\\
\caption{{ Optical images of the two target BCGs.} Left: {\textsf{i}-band SDSS DR12  image} centered at the coordinates of \BCGone. Right: \textsf{i}-band Pan-STARRS1 image centered at the coordinates of \BCGtwo. Both images are $24''\times24''$ in size. { At the bottom left of each figure a $5''$  segment is shown, along with its corresponding value in physical units, at the redshift of the BCGs.}}\label{fig:optical_images}
\end{figure*}

Some properties of the two sources are listed in Table~\ref{tab:BCG_properties}, while in Fig.~\ref{fig:optical_images} we show their optical images { taken from the data archives of the 12th release of SDSS\footnote{https://skyserver.sdss.org/dr12/en/help/docs/docshome.aspx} and  Pan-STARRS1.\footnote{https://panstarrs.stsci.edu/} The images show that the two targeted BCGs are indeed bright, with {\sf i}-band AB magnitudes of  18.0 (\BCGone) and 21.5 (\BCGtwo). A bulge-dominated morphology is also tentatively observed for 3C~244.1, { consistent with} being the most massive galaxy in the cluster.}


We  used archival  low-frequency ($<1$~GHz) radio fluxes found in the NASA/IPAC Extragalactic Database (NED) to investigate some radio properties of the two BCGs.
\BCGtwo is a steep-spectrum radio source, with $\alpha=1.6$, while \BCGone has a standard $\alpha=0.8$, typical of  optically thin synchrotron emission in the jet. Here $\alpha$ denotes the low radio frequency spectral index, where the radio spectral flux density in units of Jy at the observer frame frequency $\nu$ is expressed as $S_\nu\propto\nu^{-\alpha}$. Using the spectral indexes and low-frequency radio fluxes, the rest-frame 408~MHz luminosity densities $L_{\rm 408~MHz}=7.0\times10^{27}$~W/Hz and $1.5\times10^{25}~$W/Hz have been found for \BCGone and \BCGtwo, respectively. Therefore, \BCGone has a low-frequency radio luminosity typical of powerful FR~II sources, while \BCGtwo has a lower radio luminosity, more typical of the bulk of the radio galaxy population \citep{Zirbel1996}. {  Both our targets are powerful radio sources, which strengthens the reliability of the BCG selection by \citet{Webb2015}. Distant radio galaxies have often been found in association with BCG hosts \citep{von_der_Linden2007,Yu2018,Castignani2019,Moravec2019}.}


\section{IRAM 30m observations and data reduction}\label{sec:observations_and_data_reduction}
We observed the two BCGs using the IRAM 30m telescope at Pico Veleta in Spain. The observations of our targets were carried out 
between 3 and 6 July 2018, during the night, as part of the  observational program 065-18 (P.I.: Castignani).

We used the Eight Mixer Receiver (EMIR) to observe CO(J$\rightarrow$J-1) emission lines from the target sources at frequencies between 81 and 242~GHz, corresponding to wavelengths between 3.0 and 1.2~mm, where J is a positive integer denoting the total angular momentum. For each source the specific CO(J$\rightarrow$J-1) transitions have been chosen to maximize the likelihood of the detection, in terms of the ratio of the predicted signal to the expected rms noise. 
In particular, we targeted simultaneously  the CO(1$\rightarrow$0) and CO(2$\rightarrow$1) lines of \BCGone using the E090 and E150 receivers, respectively, while we targeted the CO(4$\rightarrow$3) line of \BCGtwo with the E230 receiver. We refer to Table~\ref{tab:BCG_properties_mol_gas} for further details.

The E090, E150, and E230 receivers can offer 4$\times$4~GHz instantaneous bandwidth covered by the correlators. Of these four bands (UI, UO, LI, LO) we used only the lower side band LI for our CO search. 
The wobbler-switching mode was used for all the observations with a frequency of 0.5~Hz and a throw of 60~arcsec { to minimize the impact of atmosphere variability.} The adopted wobbler throw is conservatively higher than the  size of our target sources, which is in fact less than a few arcsec.

We originally planned to use the Wideband Line Multiple Autocorrelator (WILMA) to cover the LI-4 GHz band in each linear polarization. The WILMA back-end gives a resolution of 2~MHz; however, the WILMA was under maintenance during the observations. Therefore, we used { only} the fast Fourier transform spectrometers (FTSs) at 200~kHz resolution to cover the  2$\times$4~GHz lower sidebands (LI and LO) for each linear polarization. For \BCGtwo we also  covered the  2$\times$4~GHz upper sidebands (UI and UO) for each linear polarization.


The targets \BCGone and \BCGtwo were observed for a total on-source observing time of 1.8~hr and 3.9~hr, respectively.
Observations were carried out in excellent weather conditions with an average precipitation water vapor (pwv) value of $\sim$3~mm;  average system temperatures $T_{\rm sys}= 106$~K and 145~K for \BCGone at 81~GHz and 161~GHz, respectively; and  $T_{\rm sys}= 204$~K for \BCGtwo at 242~GHz. 

Data reduction and analysis were performed using the {\sc CLASS} software of the {\sc GILDAS}  package\footnote{https://www.iram.fr/IRAMFR/GILDAS/}. 
We  also corrected the CO(1$\rightarrow$0) spectrum of \BCGone for a { minor} platforming level, { corresponding to 0.85~mK in Ta$^\ast$.}
The results are presented in Sect.~\ref{sec:IRAM30m_results}.

{ M51 has been used as line calibrator, while Venus has been adopted to calibrate both pointing and focus. 
The following sources, when located at an elevation similar to those of our two targets, were used as pointing calibrators: IRC~+10216, PG~1418+546, 4C~+39.25, 6C~B104451.4+715930, 3C~454.3, and 3C~345.}



\section{Results}\label{sec:results}

\subsection{Molecular gas properties}\label{sec:IRAM30m_results}
We describe in this section the results obtained with the IRAM 30m observations.
Both targets were { undetected} by our observations.
{ Based on the optical morphologies shown in Fig.~\ref{fig:optical_images} and assuming a CO-to-optical size ratio of $\sim0.5$ \citep{Young1995}, the BCGs were also unresolved } by our observations, with a beam of $\sim$10~arcsec~$\big(\frac{242~{\rm GHz}}{\nu_{\rm obs}}\big)$ at observer frame frequency $\nu_{\rm obs}$ \citep{Kramer2013}.

We removed the baseline in each spectrum by using a polynomial fit of degree one. Then we estimated rms noise levels for the antenna temperature (Ta$^\ast$)  equal to 0.18 and 0.13~mK  at 81~GHz and 161~GHz, respectively, for \BCGone, and  0.13~mK   at 242~GHz for \BCGtwo. These rms values were   estimated within the entire LI-4~GHz bandwidth and at 300~km/s resolution. We then used the rms noise levels to set 3$\sigma$ upper limits.

In Table~\ref{tab:BCG_properties_mol_gas} we list the results of our analysis, where standard efficiency corrections have been applied  to convert  Ta$^\ast$ into the main beam temperature $T_{\rm mb}$, and then  $T_{\rm mb}$ into the corresponding CO line flux, where a 5~Jy/K conversion is used. { We   adopted the following efficiency  corrections: $T_{\rm mb}/T{\rm a}^\ast = 1.17$, $1.27$, and $1.55$ for $81$, $161$, and $242$~GHz, respectively.\footnote{https://www.iram.es/IRAMES/mainWiki/Iram30mEfficiencies}}

To derive the CO(J$\rightarrow$J-1) luminosity $L^{\prime}_{\rm CO(J\rightarrow J-1)}$, in units of K~km~s$^{-1}$~pc$^2$, from the velocity integrated CO(J$\rightarrow$J-1) flux $S_{\rm CO(J\rightarrow J-1)}\,\Delta\varv\ $, in units of Jy~km~s$^{-1}$, we   used Eq.~(3) of \citet{Solomon_VandenBout2005}:
\begin{equation}
\label{eq:LpCO}
 L^{\prime}_{\rm CO(J\rightarrow J-1)}=3.25\times10^7\,S_{\rm CO(J\rightarrow J-1)}\,\Delta\varv\,\nu_{\rm obs}^{-2}\,D_L^2\,(1+z)^{-3}\,,
\end{equation}
where $\nu_{\rm obs}$ is the observer frequency in GHz of the CO(J$\rightarrow$J-1) transition, $D_L$ is the luminosity distance in Mpc, and $z$ is the redshift of the galaxy.

By assuming a Galactic CO-to-H$_2$ conversion factor $\alpha_{\rm CO}=4.36~M_\odot\,({\rm K~km~s}^{-1}~{\rm pc}^2)^{-1}$, typical of { MS} galaxies \citep{Solomon1997,Bolatto2013}, for both targets we  estimated 3$\sigma$ upper limits to the total molecular gas mass $M({\rm H_2})=\alpha_{\rm CO}L^{\prime}_{\rm CO(1\rightarrow0)}=\alpha_{\rm CO}L^{\prime}_{\rm CO(J\rightarrow J-1)}/r_{J1}$. 
Here $r_{J1}= L^{\prime}_{\rm CO(J\rightarrow J-1)}/L^{\prime}_{\rm CO(1\rightarrow0)}$ is the excitation ratio. 
We  assumed the following fiducial excitation ratios, namely $r_{21}=0.8$  \citep[typical of distant star-forming galaxies,][]{Bothwell2013,Daddi2015,Freundlich2019} and  $r_{41}=0.4$ \citep{Papadopoulos2000}. 
{ As summarized in Appendix~\ref{app:appendix}, the adopted excitation ratios and CO-to-H$_2$ conversion factors are consistent with those typically used in the literature for distant star-forming galaxies in clusters. 
Being located in the cluster cores, the target BCGs live in the densest regions of the clusters where the gas can be easily excited. As further discussed in Sect.~\ref{sec:AGNcontamination_SFR}, the use of a lower CO-to-H$_2$ conversion factor, more appropriate for star-forming galaxies above the { MS}, does not substantially impact  our final results.
We also note that CO(J$\rightarrow$J-1) lines with J$>2$ may not be optimal H$_2$ tracers since they probe denser gas than lower J$=1,2$ CO transitions. However, at increasing redshift lower J transitions become fainter and are redshifted towards lower frequencies, while higher transitions (J$>2)$  enter the millimeter domain and have  indeed been used as H$_2$ tracers in several studies (see Appendix~\ref{app:appendix}). } 


We did not attempt to set an upper limit to the continuum emission of the target galaxies by using the available total FTS 8~GHz (LI, LO) and 16~GHz (LI, LO, UI, UO) bandwidths associated with \BCGone and \BCGtwo, respectively,  for each polarization. The faintness of our targets,  the significant intrinsic atmospheric instability at millimeter wavelengths, and  the 
platforming occurring with the FTS backend (see also Sect.~\ref{sec:observations_and_data_reduction}) prevented us from determining robustly the continuum level or from estimating its upper limit.

We  used the SFRs and our molecular mass estimates to set 3$\sigma$ upper limits to the depletion timescale associated with the consumption of the molecular gas and equal to $\tau_{\rm dep}=M({\rm H_2})/{\rm SFR}$.  Similarly, we  also set 3$\sigma$ upper limits to the molecular gas-to-stellar mass ratio $M({\rm H_2})/M_\star$.  For comparison, we  computed the depletion time $\tau_{\rm dep, MS}$ and the molecular gas-to-stellar mass ratio $\big(\frac{M({\rm H_2})}{M_\star}\big)_{\rm MS}$ for MS field galaxies with redshift and stellar mass equal to those of our target galaxies, as found using empirical prescriptions by \citet{Tacconi2018}.

\begin{table*}[tb]\centering
\begin{adjustwidth}{-0.7cm}{}
\begin{center}
\begin{tabular}{ccccccccccccc}
\hline\hline
 Galaxy ID &  $z_{spec}$ & CO(J$\rightarrow$J-1)  & $\nu_{\rm obs}$ & $S_{\rm CO(J\rightarrow J-1)}$   &  $M({\rm H_2})$ & $\tau_{\rm dep}$ & $\frac{M({\rm H_2})}{M_\star}$  & $\tau_{\rm dep, MS}$ & $\big(\frac{M({\rm H_2})}{M_\star}\big)_{\rm MS}$  \\
   &  & & (GHz) &  (Jy~km~s$^{-1}$)  & ($10^{10}~M_\odot$) & ($10^8$~yr) & ($10^8~M_\odot$) & ($10^9$~yr) & \\ 
 (1) & (2) & (3) & (4) & (5) & (6) & (7) & (8) & (9) & (10)  \\
 \hline
 3C~244.1 & 0.430    & 1$\rightarrow$0 & 80.609  &  $<$0.95 & $<$4.0 & $<$1.4 & $<$0.40  &  $1.09^{+0.16}_{-0.14}$ & $0.12^{+0.19}_{-0.08}$ \\ 
     &          & 2$\rightarrow$1 & 161.215  & $<$0.77 & $<$1.0  & $<$0.36  & $<$0.10  &                   &                  \\
\hline
 SDSS~J161112.65+550823.5 & 0.907    & 4$\rightarrow$3 & 241.762  & $<$0.93 & $<$2.8  & $<$0.37  & $<$0.16  &  $0.96^{+0.18}_{-0.15}$ & $0.25^{+0.27}_{-0.13}$ \\ 
 \hline
 \end{tabular}
\end{center}
\caption{Molecular gas properties: (1) galaxy name;  (2) spectroscopic redshift as in Table~\ref{tab:BCG_properties}; (3-4) CO(J$\rightarrow$J-1) transition and observer frame frequency; (5) CO(J$\rightarrow$J-1) velocity integrated flux; (6) molecular gas mass; (7) depletion timescale  $\tau_{\rm dep}=M({\rm H_2})/{\rm SFR}$; (8) molecular gas-to-stellar mass ratio; (9-10) depletion timescale and molecular gas-to-stellar mass ratio predicted for MS field galaxies with redshift and stellar mass of our targets, following \citet{Tacconi2018}. Upper limits are at 3$\sigma$. }
\label{tab:BCG_properties_mol_gas}
\end{adjustwidth}
\end{table*}



\subsection{Comparison sample of distant (proto-)cluster galaxies}
We   compared the results found for our sources in terms of stellar mass, SFR, molecular gas content, and depletion time with those found in the literature for distant cluster galaxies with both stellar mass estimates and CO observations. In particular, we  considered the reference sample of $0.2\lesssim z\lesssim5.0$ cluster galaxies observed in CO and described in our recent work \citep[][and references therein]{Castignani2019}.

We summarize the comparison sample in the following. It includes (proto-)cluster galaxies at $z\sim0.2$ \citep{Cybulski2016}; $z\sim0.4-0.5$ \citep{Geach2011,Jablonka2013}; $z\sim1.1-1.2$ \citep{Wagg2012,Castignani2018}; $z\sim1.5-1.7$ \citep{Aravena2012,Rudnick2017,Webb2017,Noble2017,Noble2019,Hayashi2018,Kneissl2019}; $z\sim2.0-2.5$ \citep{Emonts2013,Emonts2016,Ivison2013,Tadaki2014,Dannerbauer2017,Lee2017,Coogan2018,Wang2018}; $z\sim0.4-2.6$ \citep{Castignani2019};  $z\sim3.47$ \citep[Candels-5001,][]{Ginolfi2017}; $z=4.05$ \citep{Tan2014}; $z=5.2-5.3$ \citep{Walter2012,Riechers2010}. 

We  updated this sample by replacing the results found by \citet[][]{Dannerbauer2017} and \citet[][]{Tadaki2014} for the $z\simeq2.2$ Spider~Web and $z\simeq2.5$ USS~1558-003 protocluster galaxies, respectively. In this work we
adopt the updated results by \citet{Tadaki2019}, who reported a
total of 10 CO(3$\rightarrow$2) detections for these two protoclusters.
 We  also include for the present analysis   the submillimeter galaxy J221735.15+001537.3 detected in CO(3$\rightarrow$2) by \citet{Bothwell2013} and belonging to the protocluster SSA22 at $z=3.10$ for which both stellar mass and SFR estimates are available  \citep[][]{Umehata2015}.
{ We  also considered eight additional galaxies, with stellar mass estimates, which have been recently detected in CO by \citet{GomezGuijarro2019}. Half of { the} sources have been observed in CO(3$\rightarrow$2) and belong to the HELAISS02  protocluster ($z=2.171$), while the other four have been observed in  CO(1$\rightarrow$0) and belong to the HXMM20 protocluster ($z=2.602$). }

{ This selection yields 118 sources over 33 (proto-)clusters. By also including   \BCGone and \BCGtwo the final sample comprises 120 sources, all with  $M_\star>10^{9}~M_\odot$, and over 35 (proto-)clusters at $0.2\lesssim z\lesssim5.0$.}



Including ${\rm SFR}\gtrsim3\times{\rm SFR}_{\rm MS}$ sources might result in biased-high molecular gas masses \citep[see, e.g.,][for discussion]{Noble2017,Castignani2018,Castignani2019}.  However, since our target sources have estimated ${\rm SFR}\gtrsim10\times{\rm SFR}_{\rm MS}$ we prefer not to discard such high-SFR galaxies from the comparison. 

\subsection{Distant BCGs observed in CO}
Among the  $\sim100$ distant sources with CO observations and stellar mass estimates considered in the previous section there are only a few BCG candidates. As discussed in \citet{Castignani2019} they are i) the $z=1.7$ BCG observed in CO(2$\rightarrow$1) by \citet{Webb2017}, which  belongs to the \citet{Webb2015} sample of distant BCGs, similarly to our two targets; ii) the $z = 1.99$ BCG candidate of the cluster ClJ1449+0856, detected in CO(4$\rightarrow$3) and CO(3$\rightarrow$2) by \citet{Coogan2018}; iii) MRC~1138-262 at $z = 2.2$ (i.e., the Spider~Web galaxy), detected in CO(1$\rightarrow$0) by \citet{Emonts2013, Emonts2016}; and iv) Candels-5001 at $z = 3.472$, detected in CO(4$\rightarrow$3) by \citet{Ginolfi2017}.

In addition to these four sources, in \citet{Castignani2019} we recently reported CO observations for five additional distant BCG candidates hosting radio galaxies. The sources were selected within the DES SN deep fields and COSMOS survey as part of a large search for distant $z>0.3$ star-forming radio galaxies. Our analysis yielded upper limits to the total H$_2$ gas mass for the five radio sources, namely DES~radio galaxies 399, 708, COSMOS-FR~I 16, 31, and 70 at redshifts $z=0.39$, 0.61, 0.97, 0.91, and 2.63, respectively.

By including \BCGone and SDSS~J161112.65+550823.5 from this work, our final BCG subsample consists of 11 distant BCG candidates observed in CO, spanning a broad redshift range $z\sim0.4-3.5$.
Consistently with the value adopted for \BCGone and \BCGtwo (see Sect.~\ref{sec:IRAM30m_results}) to allow a homogeneous comparison throughout this work, we assume a Galactic  CO-to-H$_2$ conversion factor $\alpha_{\rm CO}=4.36~M_\odot\,({\rm K~km~s}^{-1}~{\rm pc}^2)^{-1}$. \\

{ In Appendix~\ref{app:appendix} we summarize redshifts, stellar masses, and molecular gas properties for the compilation of 120 (proto-)cluster galaxies observed in CO, at different transitions and with different telescopes. While we adopt a constant $\alpha_{\rm CO}$ to convert velocity integrated  CO(1$\rightarrow$0) luminosities into $M({\rm H_2})$ masses, for each source we choose the excitation ratio reported in the corresponding reference study. We refer to Table~\ref{tab:CO_properties_all_galaxies} for details, while in Table~\ref{tab:CO_properties_BCG} we focus only on the 11 BCG candidates. The tables show that the sample of distant cluster galaxies observed in CO,  remarkably, has increased fourfold since 2017. For  comparison, we refer to the recent compilation by  \citet[][Table~A.1]{Dannerbauer2017} of $z>0.4$ cluster galaxies.}

{ In the cases where uncertainties for the SFR and $M_\star$ are not found in the reference studies we  assumed a fiducial 50\% error for the SFR and $\sim0.3$~dex uncertainty for $M_\star$.  Stellar mass estimates commonly rely on stellar population synthesis models and have statistical uncertainties $\sim(0.10-0.14)$~dex \citep[e.g.,][]{Roediger_Courteau2015}. An additional $\sim0.25$~dex uncertainty may be added because of the unknown initial mass function \citep[e.g.,][and references therein]{Wright2017}, resulting in a typical $\sim0.3$~dex uncertainty on $M_\star$, which is also adopted   for our two target BCGs.}


\begin{figure*}[h!]\centering
\subfloat{\hspace{0.2cm}\includegraphics[trim={0cm 0cm 4.5cm 
0cm},clip,width=0.5\textwidth,clip=true]{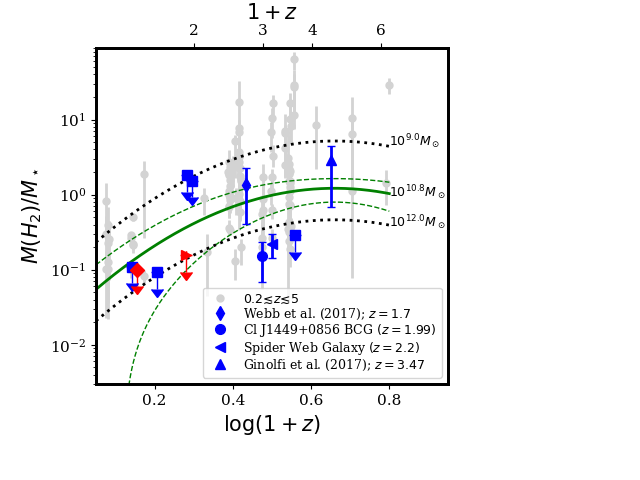}}
\subfloat{\hspace{0.2cm}\includegraphics[trim={0cm 0cm 4.5cm 
0cm},clip,width=0.5\textwidth,clip=true]{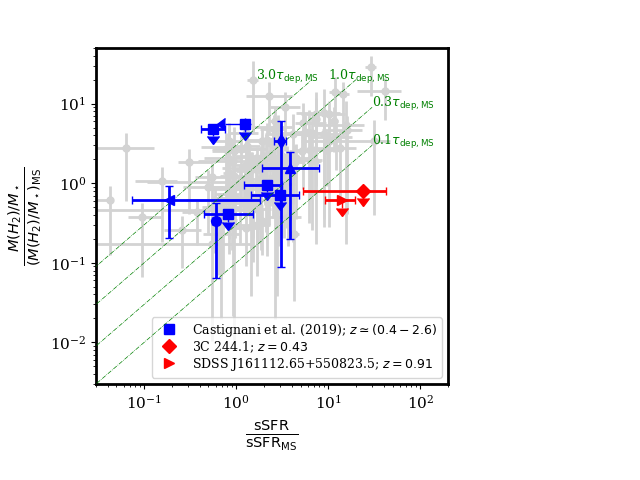}}\\
\caption{{ Molecular gas properties of distant BCGs and cluster galaxies observed in CO}. Left: Evolution of the molecular gas-to-stellar mass ratio as a function of the redshift for cluster galaxies at $0.2\lesssim z\lesssim5$ observed in CO.  
The solid green curve is the scaling relation found by \citet{Tacconi2018} for field galaxies in the MS and  with $\log(M_\star/M_\odot)$=10.8. The green dashed lines show the statistical 1$\sigma$ uncertainties in the model.  The dotted black lines show the same scaling relation as the solid green line, but for different stellar masses, $\log(M/M_\star)=$~9 and 12, which correspond to the stellar mass range spanned by the data points. Right: Molecular gas-to-stellar mass ratio vs. the specific SFR for the cluster galaxies in our sample, both normalized to the corresponding MS values using the relations { for the ratio and the SFR by \citet{Tacconi2018} and \citet{Speagle2014}, respectively}. The dot-dashed green lines show different depletion times, in units of the depletion time at the MS. { The legend for the data points is shown in the bottom right corner of each panel.}}\label{fig:mol_gas1}
\subfloat{\hspace{0.2cm}\includegraphics[trim={0.2cm 1.1cm 4.8cm 
1.2cm},clip,width=0.33\textwidth,clip=true]{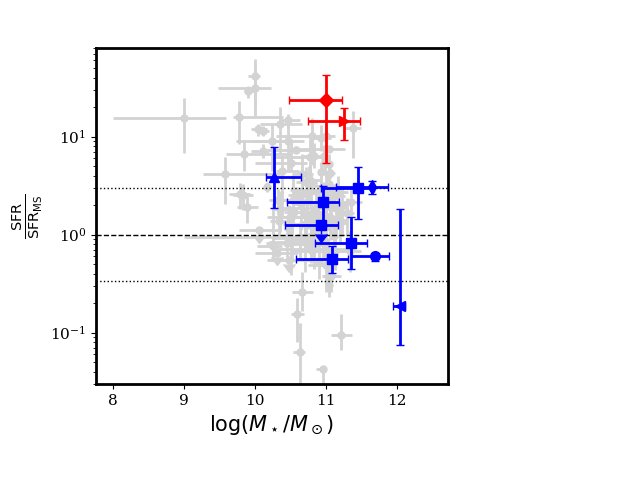}}
\subfloat{\hspace{0.2cm}\includegraphics[trim={0.2cm 1.1cm 4.8cm 
1.2cm},clip,width=0.33\textwidth,clip=true]{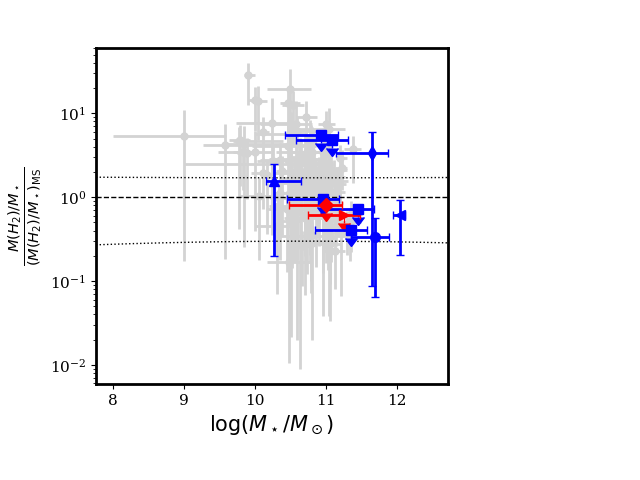}}
\subfloat{\hspace{0.2cm}\includegraphics[trim={0.2cm 1.1cm 4.8cm 
1.2cm},clip,width=0.33\textwidth,clip=true]{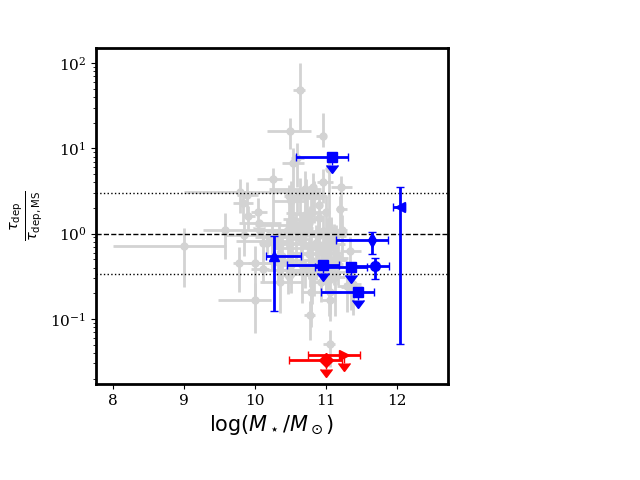}}
\caption{{ Star formation rate (left), molecular gas-to-stellar mass ratio (center), and depletion time (right), as a function of the stellar mass for distant BCGs and cluster galaxies observed in CO.} The y-axis values are all normalized to the corresponding MS values using the relations by \citet{Speagle2014} and \citet{Tacconi2018}. { The horizontal dashed lines correspond to y-axis values equal to unity, while the dotted lines denote the fiducial uncertainties associated with the MS. { The uncertainty is chosen equal to $\pm0.48$~dex for both left and right panels, since the MS is commonly identified by $1/3<{\rm SFR}/{\rm SFR}_{\rm MS}<3$. For the central panel, the plotted uncertainties are estimated at redshift $z=1$.}} The color-coding for the data points is the same as in Fig.~\ref{fig:mol_gas1}.}\label{fig:mol_gas3}
\end{figure*}

\subsection{Molecular gas, star formation, and depletion time}\label{sec:molgas}

In Fig.~\ref{fig:mol_gas1} we show the ratio of molecular gas to stellar mass  $M({\rm H}_2)/M_\star$ as a function of both redshift and specific SFR for the $\sim100$ distant sources with CO observations and stellar mass estimates. { The 11 distant BCGs} are highlighted. 
In Fig.~\ref{fig:mol_gas3} we show the SFR, $M({\rm H}_2)/M_\star$, and the depletion time ($\tau_{\rm dep})$, all normalized to their MS values, as a function of the stellar mass. 

These plots show that the BCGs indeed populate  the high-mass end of the galaxy population with CO observations, having stellar masses in the range $\log(M_\star/M_\odot)=10.3-12.0$. Given the large uncertainties and scatter in the data points, the BCGs also have  SFRs and $M({\rm H}_2)/M_\star$, both of which are  consistent with the values found using empirical relations for the MS \citep{Tacconi2018,Speagle2014} and with those of the other distant cluster galaxies observed in CO. 

However, 7 out of the 11  BCGs (i.e., the majority, $64\%\pm15\%$)\footnote{Here and throughout the text the   uncertainties on the percentage ($1\sigma$) are estimated using the binomial distribution. We refer to \citet{Castignani2014} for a similar methodology.}
have $M({\rm H}_2)/M_\star$ values that are lower than those estimated for MS galaxies of similar stellar mass and redshift by using the \citet{Tacconi2018} relation. This also applies  to our targets \BCGone and \BCGtwo. Furthermore COSMOS-FRI~16 and 31 at $z\simeq1.0$ have upper limits to the $M({\rm H}_2)/M_\star$ ratio that are higher than the MS values. Therefore, {it may also be that the fraction of BCGs with low values of $M({\rm H}_2)/M_\star$} is even higher than estimated. 
Similarly, 6 out of the 11 BCGs (i.e., the majority, $55\%\pm15\%$) have SFRs that are higher than those estimated for MS galaxies of similar stellar mass and redshift by adopting the \citet{Speagle2014} relation.

Interestingly, combining these results for the $M({\rm H}_2)/M_\star$ values and the SFR we obtain that the 11 considered BCGs have, overall, low depletion times $\tau_{\rm dep}=M({\rm H}_2)/$SFR, with estimated values or upper limits in the range (0.04-7)~Gyr. As can be seen in  Fig.~\ref{fig:mol_gas3} (right), eight BCGs (i.e., $73\%\pm18\%$) have indeed $\tau_{\rm dep}$ values that are lower than those, $\tau_{\rm dep,MS}$, estimated for the MS using the \citet{Tacconi2018} relation.
Limiting ourselves to the remaining three BCGs, the Spider~Web galaxy  has large uncertainties associated with the depletion time, while COSMOS-FRI~16 and 31 only have  upper limits to the $M({\rm H}_2)/M_\star$ ratio. These {results} suggest that the large majority or  possibly all 11 BCGs have $\tau_{\rm dep}\lesssim\tau_{\rm dep,MS}$.

{ We  also compared the properties of the 11 BCGs with those of the comparison sample by means of the Kolmogorov-Smirnov test. The null hypothesis for the $\log(M_\star/M_\odot)$ distributions of the two populations is rejected with a significance of $3.2\sigma$, { consistent} with the fact that BCGs are indeed the most massive galaxies in the { clusters}. When considering the $M_{H_2}/M_\star$, $(M_{H_2}/M_\star)/(M_{H_2}/M_\star)_{\rm MS}$, and $\tau_{\rm dep}/\tau_{\rm dep,MS}$ distributions for the two populations,  significances of $2.4\sigma$, $1.7\sigma$, and $2.0\sigma$ are found, respectively. These results suggest that there is a tentative evidence at $\sim2\sigma$ that the population of distant star-forming BCGs differ with respect to that of  distant star-forming cluster galaxies in terms of their molecular gas content and depletion time, while no statistically significant difference in terms of their SFR/SFR$_{\rm MS}$ has been found.}

Interestingly, given the high SFR (on average) associated with the 11 BCGs, they are also expected to assemble a significant fraction $f=\tau_{\rm dep}\,{\rm SFR}/M_\star$ of their stellar mass within the timescale $\sim\tau_{\rm dep}$. 
By considering upper limits to the depletion time as true measurements, we find  a median value of $f\simeq16\%$. Similarly, \citet{McDonald2014} find that a significant fraction  $f\simeq3\%$ of the stellar mass is provided
for the Phoenix~A BCG at $z=0.597$ within its starburst phase  \citep[SFR$\sim800~M_\odot/$yr,][]{McDonald2013} in a short $\tau_{\rm dep}\sim30$~Myr.
Therefore, non-negligible $f$-values and relatively short $\tau_{\rm dep}$ seem to be common for distant star-forming BCGs, quite independently of their redshifts, $z\simeq0.4-3.5$, corresponding to a 7.4~Gyr interval in cosmic time.  

{ Our results, based on observations in CO of distant $z\sim0.4-3.5$ star-forming BCGs, are consistent with a broader picture} where environmental mechanisms  (e.g., strangulation, ram pressure stripping, and galaxy harassment) tend to favor the exhaustion of the BCG gas reservoirs on a relatively short timescale $\sim\tau_{\rm dep}$.
Furthermore, while our study considers a subsample of rare star-forming BCGs  and might not be of general validity for the entire BCG population in the distant Universe, it also { suggests} a scenario where star-forming BCGs assemble a significant fraction of their stellar mass not only during  early stages ($z\gtrsim2$), but also at later epochs ($z\lesssim1$). { Our results thus support recent studies that proposed that BCGs grow by a factor of $\sim2$ in stellar mass since $z\sim1$ 
\citep{Lidman2012,Zhang2016}, but are also not in contradiction with theoretical predictions, according to which most of the BCG stellar mass is assembled at higher redshifts ($z\sim3-5$) in smaller sources which will  later be swallowed by the BCGs \citep[][]{DeLucia_Blaizot2007}.}

\subsection{AGN contamination and SFR}\label{sec:AGNcontamination_SFR}
Further considerations are needed concerning the SFR estimates of the 11 BCGs. They have $1/5\lesssim {\rm SFR/SFR}_{\rm MS}\lesssim4$, with the exception represented by the two IRAM 30m targets of this work, for which ${\rm SFR/SFR}_{\rm MS}\sim10-20$. Such  high SFR estimates might imply the need for a smaller $\alpha_{\rm CO}$ conversion factor than the Galactic value, { which we  use in this work to allow a homogeneous comparison.} A value $\alpha_{\rm CO}\simeq1~M_\odot\,({\rm K~km~s}^{-1}~{\rm pc}^2)^{-1}$ is indeed usually adopted for (ultra-)luminous infrared galaxies \citep[see, e.g.,][for a review]{Bolatto2013}.
Assuming a lower $\alpha_{\rm CO}$ than the Galactic value would imply even lower $M({\rm H}_2)/M_\star$ values for our two targets, increasing the observed tension with respect to the MS values. Alternatively,  the BCG emission is contaminated by an AGN, which might result in biased-high SFRs. In the following we consider the 11 BCGs separately.


As discussed in Sect.~\ref{sec:sample}  our IRAM 30m targets are both radio galaxies; this is also the case for the other five $z\simeq(0.4-2.6)$ sources we observed in CO with the IRAM 30m telescope \citep{Castignani2019}.  
MRC~1138-262 and the SpARCS1049+56 BCG are powerful radio sources. They have $L_{\rm 4.5~GHz}=1.9\times10^{28}$~W/Hz \citep[e.g.,][]{Carilli1997,Pentericci1997,Miley2006} and $L_{\rm 1.4~GHz}=4.2\times10^{24}$~W/Hz \citep[][]{Trudeau2019}, respectively.
\citet{Gobat2011} discussed the possibility that some obscured AGN activity is associated with the triplet of galaxies corresponding to the ClJ1449+0856 (proto-)BCG, resulting in the observed  24~$\mu$m emission.
Candels-5001 at $z=3.47$ has an X-ray (2-10~keV) luminosity of $10^{42.5}$~erg/s which \citet{Fiore2012} attributed to stellar sources. However the authors also report for the source a 1.4~GHz radio flux of $(15.3\pm6.3)~\mu$Jy, which implies a  rest-frame 1.4~GHz luminosity density $L_{\rm 1.4~GHz}\simeq1.2\times10^{24}$~W/Hz, assuming a spectral index $\alpha=0.8$ \citep[e.g.,][]{Chiaberge2009}.

Although further multiwavelength observations of the BCGs are needed to have a robust evaluation of the AGN contamination, we note that for distant star-forming galaxies it is typically $\sim20\%$ at mid- to far-infrared wavelengths \citep[e.g.,][]{Pozzi2012} and { may be higher for active BCGs.}  
{ However, \citet{Webb2015b} corrected the SFR of the SpARCS1049+56 BCG for a limited AGN contamination of $\sim20\%$. Furthermore, spectral energy distribution (SED) modeling of the five $z\simeq(0.4-2.6)$ sources of our IRAM 30m campaign, selected within the COSMOS and DES SN deep fields, shows that the infrared to ultraviolet emission  closely resembles  that of ellipticals \citep{Baldi2013,Castignani2019}, more than that,  steep-spectrum, typical of radio-loud quasars. The SEDs of these five BCGs are also consistent with those of other samples of radio loud star-forming galaxies in COSMOS \citep{Delvecchio2017}.}

\begin{figure*}[h!]\centering
\subfloat{\hspace{0.2cm}\includegraphics[trim={0.3cm 1.cm 2.4cm 1.cm},clip,width=0.5\textwidth,clip=true]{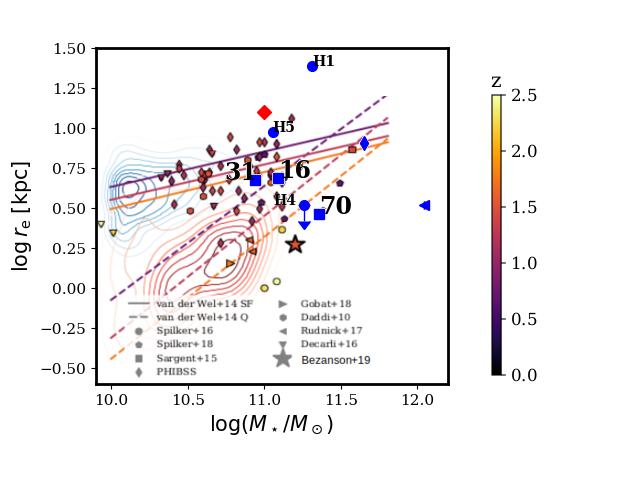}}
\subfloat{\hspace{1.2cm}\includegraphics[trim={0.5cm 1.cm 2.4cm 
1.2cm},clip,width=0.5\textwidth,clip=true]{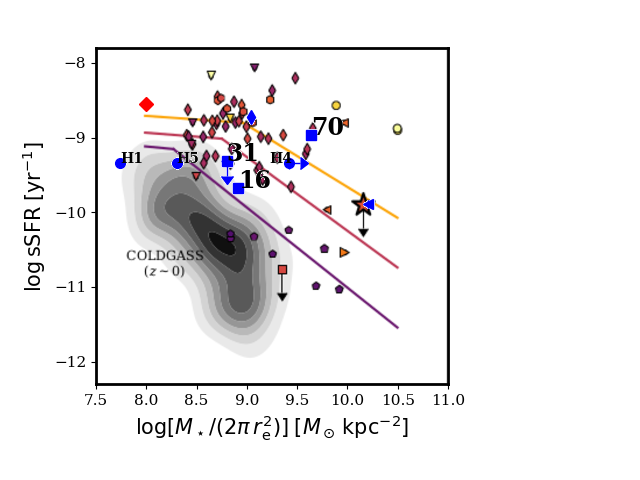}}\\
\caption{Effective radius ($r_{\rm e}$) vs. stellar mass (left) and specific SFR vs. stellar mass surface density (right). The red diamond is \BCGone, while blue symbols correspond to  ClJ1449+0856 (H1, H4, H5) and SpARCS1049+56 BCGs, MRC~1138-262, as well as COSMOS-FRI~16, 31, and 70, which are highlighted with the corresponding ID numbers. We refer to Fig.~\ref{fig:mol_gas1} for the legend. Comparison data points, contours, and empirical relations shown in Figs.~3 and 5 of \citet{Bezanson2019} are superimposed. See {text for details.}}\label{fig:reff}
\end{figure*}

\subsection{Morphological classification of the BCGs}\label{sec:reff}
{ Complex molecular gas morphologies and filamentary structures have been found associated with local gas-rich BCGs \citep[e.g.,][]{Russell2016,Russell2017}. Several observational studies have also proposed connections between the optical morphologies and the star formation properties of distant galaxies \citep{vanderWel2014,Dimauro2018,Dimauro2019,Socolovsky2018,Socolovsky2019,Puglisi2019,Freundlich2019}. Inspired by these studies we investigate the morphological properties of the 11 BCGs considered in this work, while in Sect.~\ref{sec:compactness_quenching} we discuss them in relation to their molecular gas reservoirs and star formation properties.}

We  looked for archival {\it Hubble Space Telescope} ({\it HST}) observations and found {\it HST} images for the ClJ1449+0856 ($z=1.99$) and SpARCS1049+56 ($z=1.7$) BCGs, MRC~1138-262 ($z=2.2$), and \BCGone. We did not find any archival {\it HST} image for the other IRAM 30m target \BCGtwo. In \citet{Castignani2019} we present a morphological analysis of COSMOS-FRI~16, 31, and 70, based on archival
{\it HST} ACS F814W (I-band) images with a pixel scale of 0.03~arcsec. By running {\sc Galfit} \citep{Peng2002,Peng2010} on these images we estimated half-light radii $r_{\rm e}=(4.9\pm0.8)$, $(4.7\pm0.8)$, and $(2.9\pm0.8)$~kpc for the three COSMOS-FRIs. 

For the other four BCG candidates several archival {\it HST} images have been found. When there were  multiple observations we   visually inspected the images and chose those of higher resolution and better quality, as outlined in the following.
We  considered the archival WFPC2 image (F785LP filter) at a resolution of 0.1~arcsec for 3C244.1; WFC3 images for both ClJ1449+0856 (filter F140W) and SpARCS1049+56 (filter F160W) BCGs, at a resolution of 0.128~arcsec; and the ACS/WFC image (F814W filter) for MRC~1138-262, at a resolution of 0.05~arcsec.

We   then ran {\sc Galfit} by fitting the {\it HST} images of the BCGs with a S\`{e}rsic law, as reported in Eq.~3 of \citet{Castignani2019}. The fits were performed using a generic Tiny Tim \citep{Krist1995,Krist2011} point spread function for the different {\it HST} filters { and cameras}. Our {\sc Galfit} analysis yielded $r_{\rm e}=(12.6\pm0.8)$~kpc for \BCGone,  $r_{\rm e}=(8.0\pm1.1)$~kpc for the SpARCS1049+56 BCG, and  $r_{\rm e}=(3.3\pm0.6)$~kpc for MRC~1138-262. For the ClJ1449+0856 BCG at $z\simeq2$ several components are identified in the {\it HST} images by previous studies \citep{Gobat2011,Strazzullo2018}, which suggests that the BCG is still assembling and associated with a triplet of galaxies in a likely merging phase. Similarly to these studies, we denoted the three {\it HST} components as H1, H4, and H5. Our {\sc Galfit} analysis yielded $r_{\rm e}=(24.3\pm2.3)$~kpc for H1, $r_{\rm e}=(9.5\pm1.1)$~kpc for H5, and $r_{\rm e}<3.3$~kpc for H4. For  H4 we report only a 3$\sigma$ upper limit because the $r_{\rm e}$ estimate is on the order of the pixel size (i.e., $\sim1$~kpc at the redshift of the source). 

At variance with our recent \citet{Castignani2019} study on COSMOS-FRI sources, our attempt to derive a robust S\`{e}rsic index for the BCGs did not produce reliable results. This is mainly due to the presence of several substructures for the distant Spider Web galaxy at $z\simeq2.2$, while for the other sources the resolution of $\sim0.1$~arcsec was not high enough for robust estimates of the S\`{e}rsic index.

\subsection{Compactness, { star formation, and gas depletion}}\label{sec:compactness_quenching}
The majority of the 11 considered BCGs tend to have low $M({\rm H}_2)/M_\star$ values and low depletion timescales $\tau_{\rm dep}$ {   compared   to MS galaxies and distant cluster galaxies observed in CO (see Sect.~\ref{sec:molgas}). In particular, our two target BCGs \BCGone and \BCGtwo have low ratios of molecular gas to stellar mass  $M({\rm H}_2)/M_\star\lesssim0.2$, which suggest that gas depletion in these BCGs, and in the majority of the 11 considered BCGs, was effective. Similarly, low ratios { \citep[$\lesssim10\%$,][]{Sargent2015,Gobat2018,Bezanson2019}}  have  recently been found in distant ellipticals. 
These results motivated us to use the morphological analysis outlined in the previous section for the 11 BCGs to better understand the mechanisms governing their star formation fueling.}




{ Similarly to \citet{Bezanson2019},} in Fig.~\ref{fig:reff} we show the effective radius ($r_e$) versus the stellar mass (left) and the sSFR versus the stellar mass surface density (right), defined as $\Sigma_\star=M_\star/(2\pi r_e^2)$. We report several star-forming and quiescent galaxies. 
We overplot in both panels of the figure  7  of the 11 distant BCGs that have $r_e$ estimates, as described in Sect.~\ref{sec:reff}. { For the ClJ1449+0856 BCG we  assume that the H1, H4, and H5 components have the same sSFR and contribute to the total BCG stellar mass of $M_\star=(5.0\pm2.7)\times10^{11}~M_\odot$ \citep{Gobat2011,Castignani2019} proportionally to their {\it HST} F140W-WFC3 flux.} 

{ Comparison data points, contours, and empirical relations shown in Figs.~3 and 5 of \citet{Bezanson2019} are superimposed onto our Fig.~\ref{fig:reff} as follows. In the left panel, blue and red contours show respectively the location of star-forming and quiescent galaxies at $1<z<2$ in the 3D-HST survey \citep{Brammer2011, Skelton2014,vanderWel2012}. \citet{vanderWel2014} size-mass relations at different epochs are shown as solid and dashed lines for star-forming and quiescent galaxies, respectively.  In the right panel, gray contours correspond to $z\sim0$ galaxies from the COLDGASS survey \citep{Saintonge2011}. The solid lines show the broken power-law relations at $z=0.75$, $1.25$, and $1.75$ from \citet{Whitaker2017}. In both panels the different colors of the comparison  data points and lines correspond to different redshifts, as shown in the color bar in the left panel. In both panels galaxies from several samples with molecular gas and rest-frame optical size measurements are indicated by colored symbols \citep{Daddi2010,Tacconi2010,Sargent2015,Decarli2016,Spilker2016, Spilker2018,Rudnick2017,Gobat2018,Bezanson2019}.}

As shown in the left panel of the Figure, MRC~1138-262 and COSMOS-FRI~16, 31, and 70 have effective radii that are smaller than those found for star-forming MS galaxies of similar mass and redshift using the \citet{vanderWel2014} scaling relation \citep[see also][for further discussion]{Castignani2019}. These four sources are also those where high-resolution {\it HST} images with pixel sizes $\sim(0.03-0.05)$~arcsec are available. On the other hand, for \BCGone and both ClJ1449+0856 and SpARCS1049+56 BCGs lower resolution images (pixel size $\sim0.1$~arcsec) have been used to estimate $r_e$. These sources have estimated effective radii that are consistent with those of MS field galaxies. It might be possible that the poorer resolution resulted in biased-high $r_e$ estimates for these three sources. 


Furthermore, the ClJ1449+0856 BCG is a triplet of possibly interacting sources, among them H4 has a small estimated size $r_{\rm e}<3.3$~kpc. As discussed in \citet{Strazzullo2018}, the origin of the CO emission associated with the (proto-)BCG could be  an additional star-forming $\sim300~M_\odot$/yr component of similar size as H4. These aspects show that the ClJ1449+0856 BCG system is complex, while the morphological analysis does not exclude the presence of compact star-forming components.
Similarly, MRC~1138-262 shows several clumpy substructures likely associated with stellar components, found in the archival {\it HST} ACS images (F475W and F814W filters); see also Sect.~\ref{sec:reff}.

{ Although the small sample size and the large scatter in the data points prevent us from drawing firm conclusions,} our morphological analysis reveals that five out of the seven considered BCGs (i.e., $71\%\pm17\%$) either have  smaller $r_e$ than those of  MS field star-forming galaxies or show a complex morphology with several likely star-forming components. For the remaining two BCGs (\BCGone and the SpARCS1049+56 BCG), our {\sc Galfit} analysis did not show a compact morphology. However, { it might be} that this is due to the lower resolution images (pixel size $\sim0.1$~arcsec) used for the analysis.

As a comparison, recent work by \citet{Ito2019} shows instead that distant $z\sim4$ proto-BCGs have sizes, on average, that are $\sim28\%$ larger than those of field galaxies for a fixed brightness. We suggest that the apparent discrepancy with respect to our results is ultimately due to the different sample selection. Increasing the sample of distant BCGs and CO observations will help to more accurately characterize  the population of high-$z$ BCGs.

Figure~\ref{fig:reff} (right) shows that the majority of the considered BCGs have a high sSFR and a stellar mass surface density $\Sigma_\star$ that is consistent with those of distant star-forming galaxies from the literature. However, COSMOS~FRI~70 and MRC~1138-262 have a higher $\Sigma_\star$ than the other distant BCGs, which is consistent with the fact that they are also more compact. As can be seen in the figure, high values  ($\log(\Sigma_\star/(M_\odot~{\rm kpc}^{-2}))\sim9.5-10$) have been also found for other compact sources from the literature \citep{Spilker2016,Rudnick2017,Bezanson2019}.

Interestingly, MRC~1138-262 has both $\Sigma_\star$ and sSFR values that are very similar to those reported by \citet{Bezanson2019} for their quiescent compact elliptical at $z=1.5$. However, we note that the SFR of  MRC~1138-262 is uncertain and its sSFR might by even a factor of $\sim10$ higher than reported in Fig.~\ref{fig:reff}.  For MRC~1138-262  we  assume   ${\rm SFR}=142~M_\odot$/yr \citep{Hatch2008,Emonts2016}, with large uncertainties (see Figs.~\ref{fig:mol_gas1}, \ref{fig:mol_gas3}) since an SFR of up to $\sim1400~M_\odot$/yr \citep{Seymour2012} has been reported for the BCG.



\subsection{Compactness { and quenching}}
{ The relatively low depletion times combined with the compactness associated with the distant BCGs are reminiscent of the {compaction} phase suggested for $z\sim2-4$ galaxies by  observations \citep[e.g.,][]{Barro2013,Barro2017} and by simulations \citep{Zolotov2015,Tacchella2016a,Tacchella2016b} in which massive star-forming galaxies experience an enhancement of the star formation due to gas deposition at their centers, before an inside-out gas depletion and then followed by quenching.} 

We compare our results with those independently found for different samples of distant galaxies.
Recent studies by \citet{Socolovsky2018,Socolovsky2019} show that galaxies with high sSFR$\simeq1$~Gyr$^{-1}$ in dense environments at $0.5<z<1.0$ are strongly depleted due to rapid environmental quenching, and are likely to evolve into post-starbursts (PSBs). They also tend to have larger effective radii than those in the field, possibly because the most compact star-forming galaxies are  preferentially quenched in dense environments.

These findings seem to be in apparent disagreement with the existence of compact distant star-forming BCGs (Sect.~\ref{sec:compactness_quenching}). 
The discrepancy can be possibly explained by noting that relatively compact sources  such as the Spider Web galaxies and COSMOS-FRI~16, 31, and 70  represent a rare population of BCGs 
observed during a significant star formation activity. The BCGs will deplete their molecular gas in a relatively short time and then evolve into quenched compact ellipticals,  as { the one}  reported in the recent study by \citet{Bezanson2019}.
This picture is also in agreement with recent findings by \citet{Puglisi2019}, who studied a sample of star-forming galaxies at $1.1\leq z\leq1.7$ and found a non-negligible fraction $>29\%$ of compact sources possibly observed as early PSBs.

\section{Summary and conclusions}\label{sec:summary_conclusions}
We have investigated the effect of dense Mpc-scale environments in processing molecular gas of distant star-forming  brightest cluster galaxies (BCGs) with the final goal of better understanding the processes involved in the BCG growth and the  regulation of star formation in distant BCGs. 
To this end we  observed in CO with the IRAM 30m telescope two star-forming distant BCGs: \BCGone ($z = 0.4$) and \BCGtwo ($z=0.9$) with ${\rm SFR}=281~M_\odot$/yr and 766~$M_\odot$/yr, respectively, as inferred from 24~$\mu$m {\it Spitzer}-MIPS fluxes  \citep{Webb2015}. By adopting standard CO-to-H$_2$ conversion factors we  set robust upper limits to their ratio of molecular gas to stellar mass  $M({\rm H}_2)/M_\star\lesssim0.2$ and depletion time $\tau_{\rm dep}\lesssim40$~Myr.

We  then compared these results with those found for a compilation of $\sim100$ distant cluster galaxies with CO observations and stellar mass estimates from the literature, which include nine additional BCG candidates at $z\sim0.4-3.5$, five of them are from our recent work \citep{Castignani2019} on distant radio galaxies in dense { megaparsec-scale} environments.
This comparison places the two targeted BCGs among the distant cluster galaxies with the lowest gas fractions and shortest depletion times. More in general, by considering the sample of 11 BCGs with CO observations we  found that these rare star-forming BCGs, given the large uncertainties and scatter in the data points, have  SFRs and $M({\rm H}_2)/M_\star$ that are generally consistent with the values found using empirical relations for the MS \citep{Tacconi2018,Speagle2014} and with those of the other distant cluster galaxies observed in CO. 
Nevertheless, the majority (i.e., $64\%\pm15\%$ and $73\%\pm18\%$) of the 11 BCGs have lower $M({\rm H}_2)/M_\star$ values and lower $\tau_{\rm dep}$, respectively, than those estimated for MS galaxies of similar stellar mass and redshift.
{ Similarly, the statistical analysis based on the Kolmogorov-Smirnov test tentatively suggests that the values of $M({\rm H}_2)/M_\star$ and $\tau_{\rm dep}$ for the 11 BCGs deviates, with a significance of $\sim2\sigma$, from those
of the cluster galaxies from the comparison sample.}

These results { favor} a scenario where star-forming BCGs assemble a significant fraction $f=\tau_{\rm dep}\,{\rm SFR}/M_\star\simeq16\%$ of their stellar mass within a timescale $\tau_{\rm dep}\sim(0.04-7)$~Gyr, not only during  early stages ($z\gtrsim2$), but also at later ($z\lesssim1$) epochs,  { consistent} with previous work \citep{Lidman2012,McDonald2014}.
These findings seem to favor the presence of { environmental} mechanisms (e.g., strangulation, ram pressure stripping, and galaxy harassment) that might prevent the replenishment of gas feeding the star formation, while in the meantime allowing the exhaustion of the BCG gas reservoirs on a relatively short timescale $\sim\tau_{\rm dep}$, to sustain the observed star formation. To further explore this scenario we  investigated the possibility that the compactness { may help the regulation of the star formation in distant BCGs.}

A morphological analysis  using {\sc Galfit} was done for 7  of the 11 BCGs. The seven sources, including our IRAM 30m target \BCGone,  have archival {\it HST} observations. We  found that $71\%\pm17\%$ of the BCGs are compact or show { star-forming components or substructures}. The two most compact BCGs, which are also the most distant, in the subsample of seven 
also have high stellar mass surface density $\log(\Sigma_\star/(M_\odot~{\rm kpc}^{-2}))\sim9.5-10$, similar to that previously found for other distant compact sources \citep{Spilker2016,Rudnick2017,Bezanson2019}.

{ We speculate that}, for a significant fraction of distant star-forming BCGs,  compact morphologies and star-forming components may both favor the rapid exhaustion of molecular gas and ultimately help to quench the BCGs.
Higher resolution and higher surface brightness sensitivity observations of distant BCGs will help to distinguish between different gas-processing mechanisms (e.g., galaxy harassment, strangulation, ram pressure stripping, or tidal stripping), possibly responsible for quenching the galaxies. The BCGs considered in this work are excellent targets for ALMA as well as next-generation telescopes
such as the {\it James Webb Space Telescope}.

\appendix
\section{Compilation of distant cluster galaxies}\label{app:appendix}

\onecolumn{
\begin{small}
\LTcapwidth=0.8\textwidth
\captionsetup[longtable]{margin=0.6in}
\setlength{\LTleft}{0.cm}
\begin{landscape}
\begin{longtable}{lcccccccccc} 
\hline
(proto-)cluster & galaxy & $z$ & transition & telescope & $r_{J1}$ & $\alpha_{\rm CO}$ & $L'_{\rm CO(1\rightarrow0)}$ & $M_\star$ &  SFR & ${\rm SFR}_{\rm MS}$ \\
 & & & & & & $\big(\frac{M_\odot}{{\rm K~km~s}^{-1}~{\rm pc}^2}\big)$ &  $(10^{10}~{\rm K~km~s}^{-1}~{\rm pc}^2)$  & $(10^{10}~M_\odot)$ & $(M_\odot/{\rm yr})$ & $(M_\odot/{\rm yr})$  \\
 \hline
 (1) & (2) &  (3) & (4) & (5) & (6) & (7) & (8) & (9) & (10) & (11)  \\ 
 \hline
Abell~2192 at $z=0.188$ \\ {\tiny\citep{Cybulski2016}} & J162523.6+422740 & 0.187 & $1\rightarrow0$ & LMT & 1.00 & 4.60 & $0.3^{+0.1}_{-0.1}$ & $1.8$ & $17$ & 1.9  \\ 
  & J162644.6+422530 & 0.189 & $1\rightarrow0$ & LMT & 1.00 & 4.60 & $0.4^{+0.1}_{-0.1}$ & $6.0$ & $23$ & 3.7  \\ 
  & J162508.6+423400 & 0.190 & $1\rightarrow0$ & LMT & 1.00 & 4.60 & $0.2^{+0.1}_{-0.1}$ & $10.1$ & $14$ & 4.9  \\ 
\hline
Abell~963 at $z=0.206$ \\ {\tiny\citep{Cybulski2016}} & J101727.7+384628 & 0.201 & $1\rightarrow0$ & LMT & 1.00 & 4.60 & $0.2^{+0.1}_{-0.1}$ & $7.2$ & $8$ & 4.3  \\ 
  & J101540.2+384913 & 0.204 & $1\rightarrow0$ & LMT & 1.00 & 4.60 & $0.3^{+0.1}_{-0.1}$ & $5.1$ & $3$ & 3.6  \\ 
  & J101730.0+385831 & 0.204 & $1\rightarrow0$ & LMT & 1.00 & 4.60 & $0.3^{+0.1}_{-0.1}$ & $3.0$ & $24$ & 2.7  \\ 
  & J101611.1+384924 & 0.207 & $1\rightarrow0$ & LMT & 1.00 & 4.60 & $0.3^{+0.1}_{-0.1}$ & $11.8$ & $12$ & 5.7  \\ 
  & J101618.0+390613 & 0.208 & $1\rightarrow0$ & LMT & 1.00 & 4.60 & $0.4^{+0.1}_{-0.1}$ & $6.4$ & $42$ & 4.1  \\ 
\hline
Cl~0024+16 at $z=0.395$ \\ {\tiny\citep{Geach2011}} & MIPS~J002652.5 & 0.380 & $1\rightarrow0$ & PdBI & 1.00 & 4.60 & $0.63^{+0.04}_{-0.04}$ & $9.5^{+0.4}_{-0.4}$ & $62\pm19$ & 9.8  \\ 
  & MIPS~J002703.6 & 0.396 & $1\rightarrow0$ & PdBI & 1.00 & 4.60 & $0.4^{+0.1}_{-0.1}$ & $8.7^{+0.4}_{-0.4}$ & $42\pm12$ & 9.8  \\ 
  & MIPS~J002715.0 & 0.381 & $1\rightarrow0$ & PdBI & 1.00 & 4.60 & $0.3^{+0.1}_{-0.1}$ & $11.0^{+0.5}_{-0.5}$ & $35\pm11$ & 10.7  \\ 
  & MIPS~J002621.7 & 0.380 & $1\rightarrow0$ & PdBI & 1.00 & 4.60 & $0.7^{+0.1}_{-0.1}$ & $11.2^{+0.5}_{-0.5}$ & $56\pm16$ & 10.8  \\ 
  & MIPS~J002721.0 & 0.396 & $1\rightarrow0$ & PdBI & 1.00 & 4.60 & $1.1^{+0.1}_{-0.1}$ & $9.8^{+0.5}_{-0.5}$ & $59\pm16$ & 10.5  \\ 
\hline
Cl~1416+4446 at $z=0.397$ \\ {\tiny\citep{Jablonka2013}} & GAL1416+4446 & 0.396 & $1\rightarrow0$ & PdBI & 1.00 & --- & $0.8^{+0.1}_{-0.1}$ & $16.8^{+3.1}_{-3.1}$ & $28$ & 14.4  \\ 
\hline
Cl~0926+1242 at $z=0.489$ \\ {\tiny\citep{Jablonka2013}} & GAL0926+1242-A & 0.489 & $2\rightarrow1$ & PdBI & 1.00 & --- & $0.19^{+0.03}_{-0.03}$ & $10.0^{+1.6}_{-1.6}$ & $17$ & 14.1  \\ 
  & GAL0926+1242-B & 0.489 & $2\rightarrow1$ & PdBI & 1.00 & --- & $0.16^{+0.03}_{-0.03}$ & $0.4^{+1.9}_{-0.2}$ & $8$ & 1.9  \\ 
\hline
ISCS~J1432.4+3332 at $z=1.1$ \\ {\tiny\citep{Wagg2012}} & SST24~J143235.8+333632 & 1.115 & $2\rightarrow1$ & PdBI & 1.00 & 0.80 & $1.9^{+0.4}_{-0.4}$ & $9.3^{+3.5}_{-2.6}$ & $<150$ & 47.7  \\ 
\hline
ISCS~J1426.5+3339 at $z=1.2$ \\ {\tiny\citep{Castignani2018}} & J142626.1+333827/6 & 1.163 & $2\rightarrow1$ & NOEMA & 1.00 & 4.36 & $0.25^{+0.06}_{-0.05}$ & $6.3^{+13.6}_{-4.3}$ & $28^{+12}_{-8}$ & 38.8  \\ 
\hline
XMMXCS~J2215.9-1738 at $z=1.46$ \\ {\tiny\citep{Hayashi2018}} & ALMA.01 & 1.466 & $2\rightarrow1$ & ALMA & 0.83 & 4.36 & $2.4^{+0.1}_{-0.2}$ & $8.1^{+0.8}_{-0.5}$ & $91^{+7}_{-8}$ & 65.3  \\ 
  & ALMA.02 & 1.450 & $2\rightarrow1$ & ALMA & 0.83 & 5.03 & $0.5^{+0.1}_{-0.1}$ & $3.4^{+0.7}_{-0.6}$ & $31^{+15}_{-13}$ & 34.4  \\ 
  & ALMA.03 & 1.453 & $2\rightarrow1$ & ALMA & 0.83 & 4.18 & $2.6^{+0.2}_{-0.1}$ & $11.2^{+0.3}_{-1.2}$ & $25^{+21}_{-6}$ & 81.2  \\ 
  & ALMA.04 & 1.466 & $2\rightarrow1$ & ALMA & 0.83 & 4.92 & $0.8^{+0.1}_{-0.1}$ & $3.9^{+1.0}_{-0.7}$ & $6^{+2.8}_{-2.9}$ & 38.6  \\ 
  & ALMA.05 & 1.467 & $2\rightarrow1$ & ALMA & 0.83 & 5.50 & $0.6^{+0.1}_{-0.1}$ & $2.3^{+0.7}_{-0.4}$ & $48^{+26}_{-20}$ & 26.4  \\ 
  & ALMA.06 & 1.467 & $2\rightarrow1$ & ALMA & 0.83 & 4.16 & $2.5^{+0.2}_{-0.1}$ & $12.0^{+2.4}_{-0.8}$ & $129^{+9}_{-6}$ & 86.5  \\ 
  & ALMA.07 & 1.452 & $2\rightarrow1$ & ALMA & 0.83 & 4.35 & $1.3^{+0.1}_{-0.1}$ & $8.1^{+0.2}_{-0.7}$ & $35^{+49}_{-12}$ & 64.4  \\ 
  & ALMA.08 & 1.457 & $2\rightarrow1$ & ALMA & 0.83 & 4.58 & $1.5^{+0.1}_{-0.1}$ & $5.8^{+0.3}_{-0.6}$ & $65^{+8}_{-7}$ & 50.5  \\ 
  & ALMA.09 & 1.468 & $2\rightarrow1$ & ALMA & 0.83 & 4.21 & $0.8^{+0.1}_{-0.1}$ & $10.7^{+0.2}_{-1.2}$ & $47^{+27}_{-24}$ & 79.8  \\ 
  & ALMA.10 & 1.454 & $2\rightarrow1$ & ALMA & 0.83 & 4.88 & $1.7^{+0.1}_{-0.1}$ & $4.0^{+0.3}_{-0.4}$ & $91^{+7}_{-8}$ & 38.7  \\ 
  & ALMA.11 & 1.451 & $2\rightarrow1$ & ALMA & 0.83 & 5.78 & $1.1^{+0.1}_{-0.2}$ & $1.8^{+0.6}_{-0.8}$ & $17^{+29}_{-5}$ & 22.1  \\ 
  & ALMA.12 & 1.445 & $2\rightarrow1$ & ALMA & 0.83 & 6.09 & $0.7^{+0.1}_{-0.1}$ & $1.5^{+0.2}_{-0.1}$ & $58^{+7}_{-6}$ & 19.0  \\ 
  & ALMA.13 & 1.471 & $2\rightarrow1$ & ALMA & 0.83 & 4.56 & $1.3^{+0.2}_{-0.1}$ & $6.0^{+0.4}_{-2.7}$ & $44^{+8}_{-6}$ & 53.0  \\ 
  & ALMA.14 & 1.451 & $2\rightarrow1$ & ALMA & 0.83 & 4.29 & $0.7^{+0.1}_{-0.1}$ & $9.1^{+0.2}_{-1.9}$ & $3^{+0.1}_{-2.5}$ & 69.9  \\ 
  & ALMA.15 & 1.465 & $2\rightarrow1$ & ALMA & 0.83 & 4.98 & $1.3^{+0.1}_{-0.2}$ & $3.6^{+1.7}_{-0.2}$ & $62^{+8}_{-9}$ & 36.7  \\ 
  & ALMA.16 & 1.465 & $2\rightarrow1$ & ALMA & 0.83 & 5.15 & $1.7^{+0.1}_{-0.2}$ & $3.1^{+0.3}_{-0.5}$ & $37^{+29}_{-11}$ & 32.7  \\ 
  & ALMA.17 & 1.460 & $2\rightarrow1$ & ALMA & 0.83 & 5.41 & $1.1^{+0.1}_{-0.1}$ & $2.5^{+1.8}_{-0.5}$ & $123^{+51}_{-80}$ & 27.5  \\ 
\hline
PLCK~G073.4-57.5 at $z=1.5$ \\ {\tiny\citep{Kneissl2019}} & 3 & 1.543 & $5\rightarrow4$ & ALMA & 0.32 & 4.36, 0.80 & $4.6^{+0.9}_{-0.9}$ & $3.9^{+5.7}_{-0.1}$ & $109^{+16}_{-12}$ & 41.3  \\ 
  & 8 & 1.545 & $5\rightarrow4$ & ALMA & 0.32 & 4.36, 0.80 & $0.3^{+0.1}_{-0.1}$ & $11.5^{+2.0}_{-2.4}$ & $381^{+48}_{-41}$ & 90.2  \\ 
\hline
COSMOS~cluster~candidate at $z=1.55$ \\ {\tiny\citep{Aravena2012}} & 51613 & 1.517 & $1\rightarrow0$ & JVLA & 1.00 & 3.60 & $2.4^{+0.6}_{-0.6}$ & $4.5^{+1.8}_{-1.5}$ & $114\pm17$ & 44.8  \\ 
  & 51858 & 1.556 & $1\rightarrow0$ & JVLA & 1.00 & 3.60 & $1.2^{+0.4}_{-0.4}$ & $6.0^{+1.6}_{-1.0}$ & $190\pm28$ & 57.1  \\ 
\hline
SpARCS~J022546-035517 at $z=1.59$ \\ {\tiny\citep{Noble2019}} & J0225-371 & 1.599 & $2\rightarrow1$ & ALMA & 0.77 & 4.36 & $5.3^{+0.4}_{-0.4}$ & $6.3^{+0.8}_{-0.9}$ & $173\pm76$ & 61.4  \\ 
  & J0225-460 & 1.600 & $2\rightarrow1$ & ALMA & 0.77 & 4.36 & $2.1^{+0.2}_{-0.2}$ & $9.1^{+6.0}_{-3.5}$ & $116\pm60$ & 80.2  \\ 
  & J0225-281 & 1.611 & $2\rightarrow1$ & ALMA & 0.77 & 4.36 & $3.4^{+0.3}_{-0.3}$ & $5.4^{+3.5}_{-2.6}$ & $120\pm50$ & 55.5  \\ 
  & J0225-541 & 1.611 & $2\rightarrow1$ & ALMA & 0.77 & 4.36 & $4.8^{+1.1}_{-1.1}$ & $6.6^{+0.8}_{-0.9}$ & $82\pm30$ & 64.1  \\ 
  & J0225-429 & 1.602 & $2\rightarrow1$ & ALMA & 0.77 & 4.36 & $1.1^{+0.2}_{-0.2}$ & $0.6^{+1.9}_{-0.1}$ & $178\pm83$ & 11.2  \\ 
  & J0225-407 & 1.599 & $2\rightarrow1$ & ALMA & 0.77 & 4.36 & $1.1^{+0.2}_{-0.2}$ & $0.7^{+2.6}_{-0.3}$ & $84\pm28$ & 12.5  \\ 
  & J0225-324 & 1.600 & $2\rightarrow1$ & ALMA & 0.77 & 4.36 & $0.4^{+0.1}_{-0.1}$ & $0.1^{+0.3}_{-0.1}$ & $48\pm27$ & 3.1  \\ 
  & J0225-303 & 1.596 & $2\rightarrow1$ & ALMA & 0.77 & 4.36 & $2.3^{+0.6}_{-0.6}$ & $4.4^{+0.8}_{-0.9}$ & $3\pm3$ & 47.2  \\ 
\hline
XMM-LSS~J02182-05102 at $z=1.62$ \\ {\tiny\citep{Rudnick2017}} & 30169 & 1.629 & $1\rightarrow0$ & ALMA & 1.00 & 4.36 & $0.8^{+0.1}_{-0.1}$ & $16.6^{+6.8}_{-4.8}$ & $12^{+7.5}_{-3.5}$ & 127.1  \\ 
  & 30545 & 1.624 & $1\rightarrow0$ & ALMA & 1.00 & 4.36 & $2.6^{+0.1}_{-0.1}$ & $13.8^{+5.7}_{-4.0}$ & $155.6^{+64.2}_{-45.4}$ & 110.7  \\ 
\hline
SpARCS~J033057-284300 at $z=1.63$ \\ {\tiny\citep{Noble2017}} & J0330-57 & 1.613 & $2\rightarrow1$ & ALMA & 0.77 & 4.36 & $1.4^{+0.6}_{-0.6}$ & $3.3^{+1.8}_{-1.5}$ & $36\pm21$ & 38.9  \\ 
\hline
SpARCS~J022426-032330 at $z=1.63$ \\ {\tiny\citep{Noble2017}} & J0224-3656 & 1.626 & $2\rightarrow1$ & ALMA & 0.77 & 4.36 & $1.3^{+0.3}_{-0.3}$ & $10.0^{+1.2}_{-4.4}$ & $43\pm20$ & 87.8  \\ 
  & J0224-159 & 1.635 & $2\rightarrow1$ & ALMA & 0.77 & 4.36 & $2.0^{+0.5}_{-0.5}$ & $5.9^{+2.6}_{-1.1}$ & $217\pm82$ & 60.3  \\ 
  & J0224-3680/3624 & 1.626 & $2\rightarrow1$ & ALMA & 0.77 & 4.36 & $4.7^{+0.8}_{-0.8}$ & $9.1^{+3.5}_{-1.5}$ & $68\pm24$ & 82.0  \\ 
  & J0224-396/424 & 1.634 & $2\rightarrow1$ & ALMA & 0.77 & 4.36 & $5.8^{+0.6}_{-0.6}$ & $16.2^{+3.7}_{-2.4}$ & $166\pm60$ & 125.4  \\ 
\hline
Cl~J1449+0856 at $z=1.99$ \\ {\tiny\citep{Coogan2018}} & 13 & 1.994 & $4\rightarrow3$ & ALMA/JVLA & 0.36 & 4.50 & $0.6^{+0.1}_{-0.1}$ & $2.9^{+2.9}_{-1.4}$ & $38\pm11$ & 45.7  \\ 
  & 6 & 1.983 & $1\rightarrow0$ & ALMA/JVLA & 1.00 & 4.20 & $0.7^{+0.2}_{-0.2}$ & $5.1^{+5.1}_{-2.6}$ & $118\pm12$ & 69.7  \\ 
  & N7 & 1.996 & $4\rightarrow3$ & ALMA/JVLA & 0.36 & 4.20 & $0.5^{+0.1}_{-0.1}$ & $1.2^{+1.2}_{-0.6}$ & $<26$ & 23.5  \\ 
  & B1 & 1.988 & $1\rightarrow0$ & ALMA/JVLA & 1.00 & 3.90 & $0.4^{+0.2}_{-0.2}$ & $6.5^{+6.4}_{-3.2}$ & $57\pm12$ & 83.0  \\ 
  & 3 & 1.990 & $4\rightarrow3$ & ALMA/JVLA & 0.36 & 4.70 & $0.3^{+0.1}_{-0.1}$ & $2.0^{+2.0}_{-1.0}$ & $<23$ & 35.3  \\ 
  & S7 & 1.982 & $4\rightarrow3$ & ALMA/JVLA & 0.36 & --- & $0.1^{+0.1}_{-0.1}$ & $3.0^{+3.0}_{-1.5}$ & $<26$ & 47.0  \\ 
\hline
Spider~Web~proto-cluster at $z=2.16$ \\ {\tiny\citep{Tadaki2019}} & 1138.42 & 2.163 & $3\rightarrow2$ & ALMA & 0.56 & 10.81 & $1.5^{+0.3}_{-0.3}$ & $0.6^{+0.2}_{-0.2}$ & $41\pm12$ & 15.9  \\ 
  & 1138.48 & 2.157 & $3\rightarrow2$ & ALMA & 0.56 & 5.55 & $1.8^{+0.1}_{-0.1}$ & $4.5^{+1.6}_{-1.2}$ & $144\pm66$ & 69.4  \\ 
  & 1138.54 & 2.148 & $3\rightarrow2$ & ALMA & 0.56 & 4.41 & $3.9^{+0.2}_{-0.2}$ & $15.1^{+5.3}_{-3.9}$ & $466\pm214$ & 172.1  \\ 
  & 1138.56 & 2.144 & $3\rightarrow2$ & ALMA & 0.56 & 9.64 & $1.2^{+0.2}_{-0.2}$ & $0.8^{+0.3}_{-0.2}$ & $36\pm11$ & 18.9  \\ 
\hline
USS~1558~003 at $z=2.53$ \\ {\tiny\citep{Tadaki2019}} & 1158.43 & 2.528 & $3\rightarrow2$ & ALMA & 0.56 & 6.59 & $1.1^{+0.2}_{-0.2}$ & $12.0^{+4.2}_{-3.1}$ & $66\pm21$ & 175.0  \\ 
  & 1158.54 & 2.515 & $3\rightarrow2$ & ALMA & 0.56 & 12.80 & $0.8^{+0.1}_{-0.1}$ & $2.2^{+0.8}_{-0.6}$ & $67\pm20$ & 48.6  \\ 
  & 1158.59 & 2.513 & $3\rightarrow2$ & ALMA & 0.56 & 8.69 & $1.7^{+0.1}_{-0.1}$ & $5.2^{+1.8}_{-1.4}$ & $88\pm26$ & 92.6  \\ 
  & 1158.64 & 2.529 & $3\rightarrow2$ & ALMA & 0.56 & 27.15 & $1.3^{+0.2}_{-0.2}$ & $0.7^{+0.3}_{-0.2}$ & $50\pm15$ & 19.7  \\ 
  & 1158.73 & 2.526 & $3\rightarrow2$ & ALMA & 0.56 & 13.22 & $0.6^{+0.1}_{-0.1}$ & $2.1^{+0.8}_{-0.6}$ & $106\pm31$ & 47.1  \\ 
  & 1158.137 & 2.525 & $3\rightarrow2$ & ALMA & 0.56 & 10.70 & $0.5^{+0.1}_{-0.1}$ & $3.2^{+1.1}_{-0.8}$ & $98\pm29$ & 64.5  \\ 
\hline
HELAISS02 at $z=2.17$ \\ {\tiny\citep{GomezGuijarro2019}} & S0 & 2.173 & $3\rightarrow2$ & ALMA & 0.69 & 3.50 & $11.6^{+1.1}_{-1.0}$ & $3.0^{+1.3}_{-0.7}$ & $760\pm120$ & 52.0  \\ 
  & S1 & 2.164 & $3\rightarrow2$ & ALMA & 0.69 & 9.70 & $0.7^{+0.2}_{-0.1}$ & $11.5^{+2.6}_{-2.6}$ & $269\pm37$ & 141.4  \\ 
  & S2 & 2.169 & $3\rightarrow2$ & ALMA & 0.69 & 5.60 & $2.1^{+0.2}_{-0.2}$ & $14.8^{+3.0}_{-3.0}$ & $302\pm49$ & 171.3  \\ 
  & S3 & 2.174 & $3\rightarrow2$ & ALMA & 0.69 & 6.70 & $1.0^{+0.2}_{-0.2}$ & $1.3^{+0.3}_{-0.4}$ & $200\pm32$ & 27.8  \\ 
\hline
HXMM20 at $z=2.6$ \\ {\tiny\citep{GomezGuijarro2019}} & S0 & 2.602 & $1\rightarrow0$ & VLA & 1.00 & 2.90 & $12.0^{+2.1}_{-1.8}$ & $0.8^{+0.2}_{-0.1}$ & $661\pm91$ & 22.9  \\ 
  & S1 & 2.598 & $1\rightarrow0$ & VLA & 1.00 & 3.70 & $3.5^{+1.3}_{-0.9}$ & $1.3^{+0.3}_{-0.1}$ & $380\pm44$ & 33.1  \\ 
  & S2 & 2.602 & $1\rightarrow0$ & VLA & 1.00 & 2.40 & $7.2^{+2.8}_{-2.0}$ & $1.1^{+0.4}_{-0.2}$ & $347\pm48$ & 29.2  \\ 
  & S4 & 2.597 & $1\rightarrow0$ & VLA & 1.00 & 0.80 & $20.2^{+7.8}_{-5.6}$ & $3.2^{+3.1}_{-1.7}$ & $100\pm14$ & 65.7  \\ 
\hline
HATLAS~J084933+021443 at $z=2.41$ \\ {\tiny\citep{Ivison2013}} & HATLAS~J084933~W & 2.407 & $1\rightarrow0$ & JVLA & 1.00 & 0.80 & $13.8^{+1.7}_{-1.7}$ & $24.0^{+7.6}_{-5.8}$ & $3400$ & 279.4  \\ 
  & HATLAS~J084933~T & 2.409 & $1\rightarrow0$ & JVLA & 1.00 & 0.80 & $15.7^{+2.0}_{-2.0}$ & $10.2^{+3.3}_{-2.5}$ & $1500$ & 146.4  \\ 
  & HATLAS~J084933~M & 2.418 & $1\rightarrow0$ & JVLA & 1.00 & 0.80 & $1.6^{+0.4}_{-0.4}$ & $1.0$ & $800$ & 25.4  \\ 
  & HATLAS~J084933~C & 2.414 & $1\rightarrow0$ & JVLA & 1.00 & 0.80 & $2.2^{+0.4}_{-0.4}$ & $2.3^{+2.3}_{-1.1}$ & $640$ & 47.5  \\ 
\hline
4C~23.56~proto-cluster at $z=2.49$ \\ {\tiny\citep{Lee2017}} & HAE3 & 2.486 & $3\rightarrow2$ & ALMA & 0.53 & 4.71 & $2.2^{+0.4}_{-0.4}$ & $13.0^{+1.9}_{-1.9}$ & $176\pm78$ & 182.3  \\ 
  & HAE4 & 2.478 & $3\rightarrow2$ & ALMA & 0.53 & 4.41 & $1.6^{+0.2}_{-0.2}$ & $19.7^{+5.1}_{-5.1}$ & $414\pm175$ & 248.9  \\ 
  & HAE5 & 2.487 & $3\rightarrow2$ & ALMA & 0.53 & 5.48 & $0.6^{+0.1}_{-0.1}$ & $6.1^{+1.1}_{-1.1}$ & $374\pm140$ & 102.7  \\ 
  & HAE8 & 2.486 & $3\rightarrow2$ & ALMA & 0.53 & 5.19 & $1.7^{+0.2}_{-0.2}$ & $7.8^{+2.7}_{-2.7}$ & $156\pm63$ & 123.7  \\ 
  & HAE9 & 2.486 & $3\rightarrow2$ & ALMA & 0.53 & 5.35 & $3.4^{+0.4}_{-0.4}$ & $6.8^{+3.6}_{-3.6}$ & $90\pm40$ & 111.5  \\ 
  & HAE10 & 2.486 & $3\rightarrow2$ & ALMA & 0.53 & 5.72 & $2.3^{+0.4}_{-0.4}$ & $5.1^{+2.6}_{-2.6}$ & $115\pm47$ & 89.6  \\ 
  & HAE16 & 2.483 & $3\rightarrow2$ & ALMA & 0.53 & 5.94 & $3.1^{+0.4}_{-0.4}$ & $4.4^{+4.3}_{-4.3}$ & $76\pm32$ & 80.0  \\ 
\hline
CLJ001 at $z=2.51$ \\ {\tiny\citep{Wang2018}} & 01 & 2.503 & $1\rightarrow0$ & JVLA & 1.00 & 4.06 & $2.3^{+0.2}_{-0.2}$ & $22.9^{+9.4}_{-6.7}$ & $610^{+232}_{-168}$ & 282.4  \\ 
  & 02 & 2.507 & $1\rightarrow0$ & JVLA & 1.00 & 4.06 & $1.8^{+0.3}_{-0.3}$ & $22.4^{+9.2}_{-6.5}$ & $189^{+117}_{-72}$ & 278.0  \\ 
  & 03 & 2.514 & $1\rightarrow0$ & JVLA & 1.00 & 4.08 & $0.6^{+0.1}_{-0.1}$ & $13.5^{+5.6}_{-3.9}$ & $130^{+76}_{-48}$ & 189.8  \\ 
  & 04 & 2.501 & $1\rightarrow0$ & JVLA & 1.00 & 4.08 & $0.6^{+0.1}_{-0.1}$ & $11.5^{+4.7}_{-3.4}$ & $212^{+94}_{-65}$ & 167.0  \\ 
  & 05 & 2.508 & $1\rightarrow0$ & JVLA & 1.00 & 4.08 & $2.7^{+0.4}_{-0.4}$ & $10.7^{+4.4}_{-3.1}$ & $474^{+165}_{-123}$ & 159.0  \\ 
  & 06 & 2.494 & $1\rightarrow0$ & JVLA & 1.00 & 4.09 & $4.9^{+0.4}_{-0.4}$ & $8.5^{+3.5}_{-2.5}$ & $1275^{+445}_{-330}$ & 132.6  \\ 
  & 07 & 2.505 & $1\rightarrow0$ & JVLA & 1.00 & 4.09 & $2.8^{+0.9}_{-0.9}$ & $7.9^{+3.3}_{-2.3}$ & $207^{+106}_{-70}$ & 126.4  \\ 
  & 08 & 2.513 & $1\rightarrow0$ & JVLA & 1.00 & 4.10 & $3.2^{+0.3}_{-0.3}$ & $6.8^{+2.8}_{-2.0}$ & $717^{+250}_{-185}$ & 112.3  \\ 
  & 09 & 2.500 & $1\rightarrow0$ & JVLA & 1.00 & 4.10 & $0.5^{+0.1}_{-0.1}$ & $6.3^{+2.6}_{-1.8}$ & $161^{+88}_{-57}$ & 106.0  \\ 
  & 10 & 2.506 & $1\rightarrow0$ & JVLA & 1.00 & 4.10 & $1.0^{+0.3}_{-0.3}$ & $6.3^{+2.6}_{-1.8}$ & $104^{+81}_{-45}$ & 106.2  \\ 
  & 11 & 2.506 & $1\rightarrow0$ & JVLA & 1.00 & 4.10 & $12.7^{+4.2}_{-4.2}$ & $5.4^{+2.2}_{-1.6}$ & $320^{+143}_{-99}$ & 94.0  \\ 
  & 12 & 2.515 & $1\rightarrow0$ & JVLA & 1.00 & 4.11 & $0.3^{+0.1}_{-0.1}$ & $4.7^{+1.9}_{-1.4}$ & $22^{+13}_{-8}$ & 85.0  \\ 
  & 13 & 2.505 & $1\rightarrow0$ & JVLA & 1.00 & 4.11 & $2.0^{+0.3}_{-0.3}$ & $4.7^{+1.9}_{-1.4}$ & $351^{+205}_{-130}$ & 84.6  \\ 
  & 14 & 2.515 & $1\rightarrow0$ & JVLA & 1.00 & 4.12 & $13.3^{+2.7}_{-2.7}$ & $3.5^{+1.4}_{-1.0}$ & $153^{+119}_{-67}$ & 67.7  \\ 
\hline
SSA22 at $z=3.09$ \\ {\tiny\citep{Bothwell2013,Umehata2015}} & J221735.15+001537.3 & 3.096 & $3\rightarrow2$ & PdBI & 0.52 & 1.00 & $6.3^{+2.2}_{-1.6}$ & $3.2$ & $420^{+320}_{-80}$ & 78.0  \\ 
\hline
GN20 at $z=4.05$ \\ {\tiny\citep{Carilli2010,Tan2014}} & GN20 & 4.055 & $1\rightarrow0$ & VLA & 1.00 & 1.30 & $16.2^{+4.6}_{-4.6}$ & $11.0$ & $1860\pm90$ & 250.8  \\ 
  & GN20.2a & 4.051 & $2\rightarrow1$ & EVLA & 0.84 & 2.40 & $9.2^{+5.0}_{-5.0}$ & $3.8$ & $800\pm70$ & 108.6  \\ 
  & GN20.2b & 4.056 & $2\rightarrow1$ & EVLA & 0.84 & 2.80 & $2.9^{+1.8}_{-1.8}$ & $11.0$ & $690\pm100$ & 250.8  \\ 
\hline
HDF850.1~proto-cluster at $z=5.2$ \\ {\tiny\citep{Walter2012,Serjeant_Marchetti2014}} & HDF850.1 & 5.183 & $2\rightarrow1$ & JVLA & 0.95 & 0.80 & $4.3^{+1.0}_{-1.0}$ & $13.2^{+5.7}_{-5.7}$ & $850\pm255$ & 339.2  \\ 
\hline
AzTEC-3~proto-cluster at $z=5.3$ \\ {\tiny\citep{Riechers2010}} & AzTEC-3 & 5.298 & $2\rightarrow1$ & EVLA & 0.88 & 0.80 & $6.6^{+0.9}_{-0.9}$ & $1.0^{+0.2}_{-0.2}$ & $1800$ & 43.8  \\ 
\hline
\\\\
\caption{{\bf Properties of distant $z\gtrsim0.2$ cluster galaxies observed in CO.} Notes: 
$\bullet$~\citet{Cybulski2016}  report infrared luminosities for their $z\sim0.2$ cluster galaxies, which we have converted into SFR estimates using the \citet{Kennicutt1998} relation.
$\bullet$~\citet{Jablonka2013} did not assume any value of $\alpha_{\rm CO}$; {these sources are flagged with '---' in the  $\alpha_{\rm CO}$ column.}
$\bullet$~In our previous work \citep{Castignani2018} we detected in CO(2$\rightarrow$1) two unresolved cluster galaxies, with the same redshift and $M_\star$ \citep{Zeimann2013}, that we show in the table. Consistently with \citet{Castignani2018} we also report $L'_{\rm CO(1\rightarrow0)}$ and the SFR derived from the 24~$\mu$m observer frame flux, for each of two galaxies, assuming that they equally contribute to the observed (unresolved) emission.  
$\bullet$~Consistently with the order of preference adopted by \citet{Hayashi2018}, we report the SFR estimated using both UV and 24~$\mu$m observer frame emission, when available (i.e., for the sources with ID~ALMA.01, 06, 08, 10, 12, 13, and 15). We report the SFR estimated from the {mid-IR to optical} SED for the remaining galaxies with ID:~ALMA.02, 03, 04, 05, 07, 09, 11, 14, 16, and 17). 
$\bullet$~\citet{Kneissl2019} report gas masses assuming two separate values of $\alpha_{\rm CO}$, which are listed in this Table.
$\bullet$~For the \citet{Aravena2012} sources the reported SFRs are the average between those 
derived from infrared luminosity and from SED fitting.
$\bullet$~For the  \citet{Coogan2018} Cl~J1449+0856 galaxies with $M_\star$ estimates, reported in this table, we list the SFRs derived from the 870~$\mu$m observer frame flux. In the cases where the sources are not detected in CO(1$\rightarrow$0), we have used the CO(4$\rightarrow$3) detections to estimate $L'_{\rm CO(1\rightarrow0)}$, by assuming an excitation ratio $r_{41}=0.36$, equal to the mean between the ratios  of the two sources (ID~6, B1) with both  CO(1$\rightarrow$0) and CO(4$\rightarrow$3) detections.  
{\citet{Coogan2018} did not report any value of  $\alpha_{\rm CO}$ for their source S7; this source is flagged with '---' in the  $\alpha_{\rm CO}$ column.}
$\bullet$~\citet{Wang2018}  report infrared luminosities for their $z\sim2.5$ cluster galaxies, which we have converted into SFR estimates using the \citet{Kennicutt1998} relation.
} 
\label{tab:CO_properties_all_galaxies}
\end{longtable}
\end{landscape}
\end{small}}

\onecolumn{
\begin{small}
\LTcapwidth=0.9\textwidth
\captionsetup[longtable]{margin=0.2in}
\setlength{\LTleft}{0.cm}
\begin{landscape}
\begin{longtable}{lcccccccccc} 
\hline
(proto-)cluster & galaxy & $z$ & transition & telescope & $r_{J1}$ & $\alpha_{\rm CO}$ & $L'_{\rm CO(1\rightarrow0)}$ & $M_\star$ &  SFR & ${\rm SFR}_{\rm MS}$ \\
 & & & & & & $\big(\frac{M_\odot}{{\rm K~km~s}^{-1}~{\rm pc}^2}\big)$ &  $(10^{10}~{\rm K~km~s}^{-1}~{\rm pc}^2)$  & $(10^{10}~M_\odot)$ & $(M_\odot/{\rm yr})$ & $(M_\odot/{\rm yr})$  \\
 \hline
 (1) & (2) &  (3) & (4) & (5) & (6) & (7) & (8) & (9) & (10) & (11)  \\ 
 \hline
SpARCS1049+56 at $z=1.71$ \\ {\tiny\citep{Webb2015b,Webb2017}} & J104922.6+564033 & 1.709 & $2\rightarrow1$ & LMT & 0.85 & 0.80 & $13.8^{+1.2}_{-1.2}$ & $45.0$ & $856\pm128$ & 281.6  \\ 
\hline
Cl~J1449+0856 at $z=1.99$ \\ {\tiny\citep{Gobat2011,Coogan2018}} & A1 & 1.990 & $4\rightarrow3$ & ALMA & 0.36 & 3.60 & $1.7^{+0.1}_{-0.1}$ & $50.0^{+27.0}_{-27.0}$ & $227\pm22$ & 379.2  \\ 
\hline
Spider~Web~proto-cluster at $z=2.16$ \\ {\tiny\citep{Emonts2013,Emonts2016}} \\  {\tiny \citep{Hatch2008,Hatch2009}} & MRC~1138-262 & 2.161 & $1\rightarrow0$ & ATCA & 1.00 & 4.00 & $5.6^{+1.7}_{-1.7}$ & $110.0^{+20.0}_{-20.0}$ & $142^{+1258}_{-85}$ & 765.2  \\ 
\hline
Candels-5001 at $z=3.473$ \\ {\tiny\citep{Ginolfi2017}} & Candels-5001 & 3.473 & $4\rightarrow3$ & ALMA & 0.33 & 10.00 & $1.2^{+0.6}_{-0.5}$ & $1.9^{+2.6}_{-0.4}$ & $214^{+223}_{-109}$ & 55.6  \\ 
\hline
DES-RG~399 at $z=0.3$ \\ {\tiny\citep{Castignani2019}} & DES-RG~399 & 0.388 & $3\rightarrow2$ & IRAM~30m & 0.55 & 4.36 & $<0.2$ & $9.1$ & $21^{+10}_{-9}$ & 9.8  \\ 
\hline
DES-RG~708 at $z=0.6$ \\ {\tiny\citep{Castignani2019}} & DES-RG~708 & 0.606 & $3\rightarrow2$ & IRAM~30m & 0.55 & 4.36 & $<0.6$ & $28.2$ & $112^{+68}_{-58}$ & 37.0  \\ 
\hline
COSMOS-FRI~16 at $z=0.9$ \\ {\tiny\citep{Castignani2019}} & COSMOS-FRI~16 & 0.969 & $4\rightarrow3$ & IRAM~30m & 0.40 & 4.36 & $<4.3$ & $12.3$ & $26^{+9}_{-7}$ & 46.2  \\ 
\hline
COSMOS-FRI~31 at $z=0.9$ \\ {\tiny\citep{Castignani2019}} & COSMOS-FRI~31 & 0.912 & $4\rightarrow3$ & IRAM~30m & 0.40 & 4.36 & $<3.6$ & $8.7$ & $<42$ & 33.3  \\ 
\hline
COSMOS-FRI~70 at $z=2.6$ \\ {\tiny\citep{Castignani2019}} & COSMOS-FRI~70 & 2.625 & $7\rightarrow6$ & IRAM~30m & 0.39 & 4.36 & $<1.5$ & $22.9$ & $245^{+205}_{-112}$ & 298.0  \\ 
\hline
SpARCS~J1033+58 at $z=0.43$ \\ {\tiny{ \bf(This~work)}} & 3C~244.1 & 0.430 & $2\rightarrow1$ & IRAM~30m & 0.80 & 4.36 & $<0.2$ & $10.0$ & $281\pm218$ & 11.8  \\ 
\hline
SpARCS~J1611+55 at $z=0.91$ \\ {\tiny{ \bf(This~work)}} & SDSS~J161112.65+550823.5 & 0.907 & $4\rightarrow3$ & IRAM~30m & 0.40 & 4.36 & $<0.6$ & $18.0$ & $766\pm275$ & 53.7  \\ 
\hline
\\\\
\caption{{\bf Properties of distant BCG (candidates) observed in CO.} Notes: 
$\bullet$~For the Cl~J1449+0856 BCG the reported SFR is derived from the 870~$\mu$m observer frame flux \citep{Coogan2018}; to estimate $L'_{\rm CO(1\rightarrow0)}$, an excitation ratio $r_{41}=0.36$ is assumed, equal to the mean between the ratios  of the two \citet{Coogan2018} sources (ID~6, B1) with both  CO(1$\rightarrow$0) and CO(4$\rightarrow$3) detections.  
The reported values for  
$\alpha_{\rm CO}$ and $L'_{\rm CO(1\rightarrow0)}$ imply a molecular gas mass $\sim10^{10.8}~M_\odot$, while \citet{Coogan2018} find a dynamical mass equal to  $10^{10.3\pm0.3}~M_\odot$. The two estimates are fairly consistent with each other within the uncertainties.
$\bullet$~For MRC~1138-262 we report the SFR$=142~M_\odot$/yr \citep{Emonts2016}, with uncertainties derived assuming maximum  and minimum values equal to SFR$ = 1400 M_\odot$/yr \citep{Emonts2013} and SFR$=57~M_\odot$/yr \citep{Hatch2008}, respectively.
}
\label{tab:CO_properties_BCG}
\end{longtable}
\end{landscape}
\end{small}}

\twocolumn{
\begin{acknowledgements}
{ We thank the anonymous referee for helpful comments which contributed to improve the paper significantly.}
GC acknowledges financial support from the Swiss National Science Foundation (SNSF).
This work is based on observations carried out under project number 065-18 with the IRAM 30m telescope. IRAM is supported by INSU/CNRS (France), MPG (Germany) and IGN (Spain). This publications has made use of data products from the NASA/IPAC Extragalactic Database (NED).
\end{acknowledgements}}


\begin{thebibliography}{}
\bibitem[Alberts et al.(2016)]{Alberts2016}  Alberts, S., Pope, A., Brodwin, M. et al. 2016, ApJ, 825, 72
\bibitem[Aravena et al.(2012)]{Aravena2012} Aravena, M., Carilli, C.~L., Salvato, M. et al. 2012, MNRAS, 426, 258 
\bibitem[Bai et al.(2009)]{Bai2009} { Bai, L., Rieke, G.~H., Rieke, M.~J. et al. 2009, ApJ, 693, 1840  }
\bibitem[Baldi et al.(2013)]{Baldi2013} { Baldi, T., Chiaberge, M., Capetti, A.  et al., 2013, ApJ, 762, 30}
\bibitem[Barro et al.(2013)]{Barro2013} { Barro, G., Faber, S.~M., P\'{e}rez-Gonz\'{a}lez, P.~G. et al. 2013, ApJ, 765, 104}
\bibitem[Barro et al.(2017)]{Barro2017} Barro, G., Faber, S.~M., Koo, D.~C. et al. 2017, ApJ, 840, 47
\bibitem[Bell et al.(2003)]{Bell2003} { Bell, E.~F., McIntosh, D.~H., Katz, N. et al., 2003, ApJS, 149, 289}
\bibitem[Bezanson et al.(2019)]{Bezanson2019} { Bezanson, R., Spilker, J.,  Williams, C.~C. et al. 2019, ApJ, 873, 19}
\bibitem[Bigiel et al.(2008)]{Bigiel2008} Bigiel F., Leroy A., Walter F. et al. 2008, AJ, 136, 2846
\bibitem[Bleem et al.(2015)]{Bleem2015} { Bleem, L.~E., Stalder, B., de Haan, T. et al. 2015, ApJS, 216, 27}
\bibitem[Bolatto et al.(2013)]{Bolatto2013} Bolatto, A.~T. et al. 2013, ARA\&A, 51, 207
\bibitem[Bonaventura et al.(2017)]{Bonaventura2017} { Bonaventura, N.~R., Webb, T.~M.~A.,  Muzzin, A. et al. 2017, MNRAS, 469, 1259}
\bibitem[Bothwell et al.(2013)]{Bothwell2013} Bothwell, M. S., Smail, I., Chapman, S. C., et al. 2013, MNRAS, 429, 3047
\bibitem[Brammer et al.(2011)]{Brammer2011} Brammer, G.~B., Whitaker, K.~E., van Dokkum, P.~G., et al. 2011, ApJ, 739, 24
\bibitem[Brodwin et al.(2013)]{Brodwin2013} { Brodwin, M., Stanford, S.~A., Gonzalez, A.~H. et al.   2013, ApJ 779, 138}
\bibitem[Bruzual \& Charlot(2003)]{Bruzual_Charlot2003} Bruzual, G. \& Charlot, S., 2003, MNRAS, 344, 1000
\bibitem[Carilli et al.(1997)]{Carilli1997} { Carilli, C.~L., R\"{o}ttgering, H.~J.~A., van Ojik, R. et al. 1997, ApJS, 109, 1 }
\bibitem[Carilli et al.(2010)]{Carilli2010} Carilli, C. L., Daddi, E., Riechers, D. et al. 2010, ApJ, 714, 1407
\bibitem[Carilli \& Walter(2013)]{Carilli_Walter2013} Carilli, C.~L., \& Walter, F. 2013, ARA\&A, 51, 105
\bibitem[Castignani et al.(2014)]{Castignani2014} { Castignani, G., Chiaberge, M., Celotti. A., et al. 2014, ApJ, 792, 114}
\bibitem[Castignani et al.(2018)]{Castignani2018} { Castignani, G., Combes, F., Salom\'{e}, P. et al. 2018,  A\&A, 617, 103}
\bibitem[Castignani et al.(2019)]{Castignani2019} { Castignani, G., Combes, F., Salom\'{e}, P. et al. 2019, A\&A, 623, 48}
\bibitem[Chary \& Elbaz(2001)]{Chary_Elbaz2001} Chary, R. \& Elbaz, D. 2001, ApJ,  556, 562
\bibitem[Chiaberge et al.(2009)]{Chiaberge2009} Chiaberge, M., Tremblay, G., Capetti, A., et al. 2009, ApJ, 696, 1103
\bibitem[Chung et al.(2010)]{Chung2010} { Chung, S.~M., Gonzalez, A.~H., Clowe, D. et al. 2010, ApJ, 725, 153}
\bibitem[Collins et al.(2009)]{Collins2009} { Collins, C.~A., Stott, J.~P., Hilton, M. et al. 2009, Nature, 458, 603}
\bibitem[Coogan et al.(2018)]{Coogan2018} { Coogan, R.~T., Daddi, E., Sargent, M.~T. et al. 2018, MNRAS, 479, 703}
\bibitem[Cooke et al.(2016)]{Cooke2016} { Cooke, E.~A., Hatch, N.~A., Stern, D. et al. 2016, ApJ, 816, 83}
\bibitem[Cybulski et al.(2016)]{Cybulski2016} { Cybulski, R., Yun, M.~S., Erickson, N. et al. 2016, MNRAS, 459, 3287}
\bibitem[Daddi et al.(2010)]{Daddi2010} Daddi, E., Bournaud, F., Walter, F., et al. 2010, ApJ, 713, 686
\bibitem[Daddi et al.(2015)]{Daddi2015} Daddi, E., Dannerbauer, H., Liu, D. et al. 2015, A\&A, 577, 46
\bibitem[Dannerbauer et al.(2017)]{Dannerbauer2017} { Dannerbauer, H., Lehnert, M.~D., Emonts, B. et al. 2017, A\&A, 608, 48}
\bibitem[Decarli et al.(2016)]{Decarli2016} Decarli, R., Walter, F., Aravena, M., et al. 2016, ApJ, 833, 70
\bibitem[Dekel et al.(2009a)]{Dekel2009a} { Dekel, A., Birnboim, Y., Engel, G. et al. 2009a, Nature, 457, 451}
\bibitem[Dekel et al.(2009b)]{Dekel2009b} { Dekel, A., Sari, R., Ceverino, D. et al. 2009b, ApJ, 703, 785}
\bibitem[De Lucia \& Blaizot(2007)]{DeLucia_Blaizot2007} De Lucia, G., \& Blaizot, J. 2007, MNRAS, 375, 2
\bibitem[Delvecchio et al.(2017)]{Delvecchio2017} Delvecchio, I., Smolcic, V., Zamorani, G. et al. 2017, A\&A, 602, 3
\bibitem[Demarco et al.(2010)]{Demarco2010} { Demarco, R., Wilson, G., \& Muzzin, A. et al. 2010, ApJ, 711, 1185}
\bibitem[Dimauro et al.(2018)]{Dimauro2018} Dimauro, P., Huertas-Company, M., Daddi, E. et al. 2018, MNRAS, 478, 5410
\bibitem[Dimauro et al.(2019)]{Dimauro2019} Dimauro, P., Huertas-Company, M., Daddi, E. et al. 2019, MNRAS, 489, 4135
\bibitem[Dressler et al.(1980)]{Dressler1980} Dressler, A. 1980, ApJ, 236, 351
\bibitem[Edge(2001)]{Edge2001} Edge, A. C., 2001, MNRAS, 328, 762
\bibitem[Emonts et al.(2013)]{Emonts2013} { Emonts, B.~H.~C.,  Feain, I., R\"{o}ttgering, H.~J.~A. et al. 2013, MNRAS, 430, 3465}
\bibitem[Emonts et al.(2016)]{Emonts2016} { Emonts, B.~H.~C., Lehnert, M.~D., Villar-Mart\'{i}n, M. et al. 2016, Science, 354, 1128}
\bibitem[Fanaroff \& Riley(1974)]{Fanaroff_Riley1974} Fanaroff, B. L., \& Riley, J. M. 1974, MNRAS, 167, 31P
\bibitem[Fiore et al.(2012)]{Fiore2012} { Fiore, F., Puccetti, S., Grazian, A. et al. 2012, A\&A, 537, 16}
\bibitem[Fogarty et al.(2019)]{Fogarty2019} { Fogarty, K., Postman, M., Li, Y. et al. 2019, ApJ, 879, 103}
\bibitem[Fraser-McKelvie et al.(2014)]{FraserMcKelvie2014} Fraser-McKelvie, A., Brown, M.~J.~I., \& Pimbblet, K.~A., 2014, MNRAS, 444, 63
\bibitem[Freundlich et al.(2019)]{Freundlich2019} Freundlich, J., Combes, F., Tacconi, L. J. et al., 2019, A\&A, 622, 105
\bibitem[Geach et al.(2011)]{Geach2011} { Geach, J.~E., Smail, I., Moran, S.~M. et al. 2011, ApJL, 730, 19}
\bibitem[Ginolfi et al.(2017)]{Ginolfi2017} { Ginolfi, M., Maiolino, R., Nagao, T  et al. 2017,  MNRAS, 468, 3468}
\bibitem[Gobat et al.(2011)]{Gobat2011} {  Gobat, R., Daddi, E., Onodera, M. et al. 2011, A\&A, 526, 133}
\bibitem[Gobat et al.(2018)]{Gobat2018} { Gobat, R., Daddi, E., Magdis, G., et al. 2018, Nature Astronomy, 2, 239}
\bibitem[G\'{o}mez-Guijarro et al.(2019)]{GomezGuijarro2019} G\'{o}mez-Guijarro, C., Riechers, D.~A., Pavesi, R. et al. 2019, ApJ, 872, 117
\bibitem[Gunn \& Gott(1972)]{Gunn_Gott1972} Gunn, J.~E. \& Gott, J.~R.~III, 1972, ApJ, 176, 1
\bibitem[Hamer et al.(2012)]{Hamer2012} { Hamer, S.~L.,  Edge, A.~C., Swinbank, A.~M. et al. 2012, MNRAS, 421, 3409} 
\bibitem[Hatch et al.(2008)]{Hatch2008} { Hatch, N.~A., Overzier, R.~A., R\"{o}ttgering, H.~J.~A. et al. 2008, MNRAS, 383, 931}
\bibitem[Hatch et al.(2009)]{Hatch2009} { Hatch, N.~A., Overzier, R.~A., Kurk, J.~D. et al. 2009,  MNRAS, 395, 114}
\bibitem[Hausman \& Ostriker(1978)]{Hausman_Ostriker1978} { Hausman, M.~A. \& Ostriker, J.~P., 1978. , ApJ, 224, 320}
\bibitem[Hayashi et al.(2018)]{Hayashi2018} { Hayashi, M., Tadaki, K., Kodama, T. et al. 2018, ApJ, 856, 118}
\bibitem[Hill \& Lilly(1991)]{Hill_Lilly1991} Hill, G.~J., \& Lilly, S.~J., 1991, ApJ, 367, 1
\bibitem[Ito et al.(2019)]{Ito2019}  { Ito, K., Kashikawa, N., Toshikawa, J. et al. 2019, ApJ, 878, 68}
\bibitem[Ivison et al.(2013)]{Ivison2013} { Ivison, R.~J., Swinbank, A.~M., Smail, I. et al. 2013, ApJ, 772, 137}
\bibitem[Jablonka et al.(2013)]{Jablonka2013} { Jablonka, P., Combes, F., Rines, K. et al. 2013, A\&A, 557, 103}
\bibitem[Kramer et al.(2013)]{Kramer2013} { Kramer, C.,  Pe\~{n}alver, J. \& Greve A., 2013, \emph{Improvement of the IRAM 30m telescope beam pattern}, www.iram-institute.org$/$medias$/$uploads$/$eb2013-v8.2.pdf}
\bibitem[Kennicutt(1998)]{Kennicutt1998} Kennicutt, R.~C.~J. 1998, ARAA, 36, 189
\bibitem[Kneissl et al.(2019)]{Kneissl2019} { Kneissl, R., del Carmen Polletta, M., Martinache, C. et al. 2019, A\&A, 625, 96}
\bibitem[Kodama et al.(2001)]{Kodama2001} { Kodama, T., Smail, I., Nakata, F. et al., 2001, ApJ, 562, 9}
\bibitem[Koyama et al.(2014)]{Koyama2014} {  Koyama, Y., Kodama, T., Tadaki, K. et al. 2014, 789, 18}
\bibitem[Krist(1995)]{Krist1995} Krist, J. 1995, ASPC, 77, 349
\bibitem[Krist et al.(2011)]{Krist2011} Krist, J. E., Hook, R. N. \& Stoehr, F. 2011, SPIE, 8127
\bibitem[Larson et al.(1980)]{Larson1980} { Larson, R.~B., Tinsley, B.~M., Caldwell, C.~N. et al. 1980, ApJ, 237, 69}
\bibitem[Lauer et al.(2014)]{Lauer2014} { Lauer, T.~R., Postman, M., Strauss, M.~A. et al., 2014, ApJ, 797, 82}
\bibitem[Lee et al.(2017)]{Lee2017} { Lee, M.~M., Tanaka, I., Kawabe, R. et al. 2017,  ApJ, 842, 55}
\bibitem[Leroy et al.(2013)]{Leroy2013} Leroy A.~K., Walter F., Sandstrom K. et al. 2013, AJ, 146, 19
\bibitem[Lidman et al.(2012)]{Lidman2012} { Lidman, C., Suherli, J., Muzzin, A. et al., 2012, MNRAS, 427, 550}
\bibitem[Madau \& Dickinson(2014)]{Madau_Dickinson2014} Madau P. \& Dickinson M. 2014, ARA\&A, 52, 415
\bibitem[McDonald et al.(2013)]{McDonald2013} McDonald, M., Benson, B., Veilleux, S. et al. 2013, ApJL, 765, 37
\bibitem[McDonald et al.(2014)]{McDonald2014} McDonald, M., Swinbank, M., Edge, A.~C. et al. 2014, ApJ, 784, 18
\bibitem[McDonald et al.(2016)]{McDonald2016} McDonald, M., Stalder, B., Bayliss, M., et al. 2016, ApJ, 817, 86
\bibitem[McNamara et al.(2014)]{McNamara2014} {  McNamara, B.~R., Russell, H.~R., Nulsen, P.~E.~J. et al. 2014, ApJ, 785, 44} 
\bibitem[Miley et al.(2006)]{Miley2006} { Miley, G.~K., Overzier, R.~A., Zirm, A.~W. et al. 2006, ApJ, 650, 29} 
\bibitem[Moore et al.(1999)]{Moore1999} { Moore, B.,  Lake, G., Quinn, T. et al. 1999, MNRAS, 304, 465}
\bibitem[Moravec et al.(2019)]{Moravec2019} { Moravec E., Gonzalez A.~H., Stern D. et al. 2019, ApJ, 871, 186}
\bibitem[Muzzin et al.(2009)]{Muzzin2009} Muzzin, A., Wilson, G., \& Yee, H.~K.~C. 2009, ApJ, 698, 1934
\bibitem[Muzzin et al.(2012)]{Muzzin2012} Muzzin, A., Wilson, G., \& Yee, H.~K.~C. 2012, ApJ, 746, 188
\bibitem[Newman et al.(2014)]{Newman2014} { Newman, A.~B., Ellis, R.~S., Andreon, S. et al. 2014, ApJ, 788, 51}
\bibitem[Noble et al.(2017)]{Noble2017} Noble, A.~G., McDonald, M., Muzzin, A., et al. 2017, ApJ, 842, 21
\bibitem[Noble et al.(2019)]{Noble2019} { Noble, A.~G., Muzzin, A., McDonald, M. et al. 2019, ApJ, 870, 56}
\bibitem[Ocvirk et al.(2008)]{Ocvirk2008} { Ocvirk, P., Pichon, C., Teyssier, R. et al.  2008, MNRAS, 390, 1326}
\bibitem[Olivares et al.(2019)]{Olivares2019} {  Olivares, V., Salom\'{e}, P., Combes, F., et al. 2019, A\&A, 631, 22O }
\bibitem[Ostriker \& Tremaine(1975)]{Ostriker_Tremaine1975} Ostriker, J.~P. \& Tremaine, S.~D., 1975, ApJ, 202, 113
\bibitem[Papadopoulos et al.(2000)]{Papadopoulos2000} { Papadopoulos, P.~P., R\"{o}ttgering, H.~J.~A., van der Werf, P.~P. et al. 2000, ApJ, 528, 626}
\bibitem[Peng et al.(2002)]{Peng2002} { Peng, C.~Y., Ho, L.~C., Impey, C.~D. et al. 2002, AJ, 124, 266}
\bibitem[Peng et al.(2010)]{Peng2010} { Peng, Y-J, Lilly, S.~J., Kova\v{c}, K. et al.,  2010, ApJ, 721, 193}
\bibitem[Pentericci et al.(1997)]{Pentericci1997} { Pentericci, L., R\"{o}ttgering, H.~J.~A.,  Miley, G.~K. et al. 1997, A\&A, 326, 580}
\bibitem[Planck Collaboration(2018)]{PlanckCollaborationVI2018} Planck Collaboration results VI, 2018, arXiv:180706209
\bibitem[Pozzi et al.(2012)]{Pozzi2012} { Pozzi, F., Vignali, C., Gruppioni, C. et al. 2012, MNRAS, 423 1909}
\bibitem[Puglisi et al.(2019)]{Puglisi2019} { Puglisi, A., Daddi, E., Liu, D. et al. 2019, ApJ, 877, 23}
\bibitem[Riechers et al.(2010)]{Riechers2010} { Riechers, D.~A.; Capak, P.~L., Carilli, C.~L. et al. 2010, ApJ, 720, 131}
\bibitem[Riess et al.(2019)]{Riess2019} { Riess, A.~G., Casertano, S., Yuan, W. et al. 2019, ApJ, 876, 85}
\bibitem[Roediger \& Courteau(2015)]{Roediger_Courteau2015} Roediger, J.~C. \& Courteau, S., 2015, MNRAS, 452, 3209 
\bibitem[Rudnick et al.(2017)]{Rudnick2017} Rudnick, G., Hodge, J., Walter, F. et al. 2017, ApJ, 849, 27
\bibitem[Russell et al.(2014)]{Russell2014} { Russell, H.~R., McNamara, B.~R.,  Edge, A.~C. et al. 2014,  ApJ, 784, 78} 
\bibitem[Russell et al.(2016)]{Russell2016} Russell, H.~R., McNamara, B.~R., Fabian, A.~C. et al. 2016, MNRAS, 458, 3134
\bibitem[Russell et al.(2017)]{Russell2017} Russell, H.~R., McDonald, M., McNamara, B.~R.  et al. 2017, ApJ,  836, 130
\bibitem[Russell et al.(2019)]{Russell2019} { Russell, H.~R., McNamara, B.~R., Fabian, A.~C. et al. 2019, MNRAS, 490, 3025}
\bibitem[Saintonge et al.(2011)]{Saintonge2011} Saintonge, A., Kauffmann, G., Kramer, C., et al. 2011, MNRAS, 415, 32
\bibitem[Salom\'{e} \& Combes(2003)]{Salome_Combes2003} Salom\'{e} P. \&  Combes, F. 2003, A\&A, 412, 657
\bibitem[Salom\'{e} et al.(2006)]{Salome2006} { Salom\'{e} P., Combes, F., Edge, A.~C. et al. 2006, A\&A, 454, 437}
\bibitem[Santos et al.(2015)]{Santos2015} { Santos, J.~S., Altieri, B., Valtchanov, I. et al. 2015, MNRAS, 447, 65}
\bibitem[Sargent et al.(2015)]{Sargent2015} Sargent, M. T., Daddi, E., Bournaud, F., et al. 2015, ApJL, 806, L20
\bibitem[Schruba et al.(2011)]{Schruba2011} Schruba A., Leroy A.~K., Walter F. et al. 2011, AJ, 142, 37  
\bibitem[Serjeant \& Marchetti(2014)]{Serjeant_Marchetti2014} Serjeant, S. \& Marchetti, L. 2014, MNRAS, 443, 3118
\bibitem[Seymour et al.(2012)]{Seymour2012} { Seymour, N., Altieri, B., De Breuck, C. et al. 2012, ApJ, 755, 146}
\bibitem[Skelton et al.(2014)]{Skelton2014} Skelton, R. E., Whitaker, K. E., Momcheva, I. G., et al. 2014, ApJS, 214, 24
\bibitem[Smith et al.(2010)]{Smith2010} { Smith, G.~P., Haines, C.~P., Pereira, M.~J. et al. 2010, A\&A, 518, 1}
\bibitem[Socolovsky et al.(2018)]{Socolovsky2018} Socolovsky, M., Almaini, O., Hatch, N.~A. et al. 2018, MNRAS, 476, 1242
\bibitem[Socolovsky et al.(2019)]{Socolovsky2019} Socolovsky, M., Maltby, D.~T., Hatch, N.~A. et al.  2019, MNRAS, 482, 1640
\bibitem[Solomon et al.(1997)]{Solomon1997} { Solomon, P.~M., Downes, D., Radford, S.~J.~E. et al. 1997, ApJ, 478, 144}
\bibitem[Solomon \& Vanden Bout(2005)]{Solomon_VandenBout2005} Solomon, P.~M. \& Vanden Bout, P.~A., 2005, ARA\&A, 43, 677
\bibitem[Speagle et al.(2014)]{Speagle2014} { Speagle, J.~S., Steinhardt, C.~L., Capak, P.~L. et al. 2014, ApJS, 214, 15}
\bibitem[Spilker et al.(2018)]{Spilker2018} { Spilker, J., Bezanson, R., Bari\v{s}i\'{c}, I. et al. 2018, ApJ, 860, 103}
\bibitem[Spilker et al.(2016)]{Spilker2016} Spilker, J.~S., Bezanson, R., Marrone, D.~P., et al. 2016, ApJ, 832, 19
\bibitem[Strazzullo et al.(2016)]{Strazzullo2016} { Strazzullo, V., Daddi, E., Gobat, R. et al. 2016, ApJ, 833, 20}
\bibitem[Strazzullo et al.(2018)]{Strazzullo2018} { Strazzullo, V.; Coogan, R.~T., Daddi, E. et al. 2018, ApJ, 862, 64}
\bibitem[Strazzullo et al.(2019)]{Strazzullo2019} { Strazzullo, V., Strazzullo, V., Pannella, M., Mohr, J.~J. et al. 2019, A\&A, 622, 117}
\bibitem[Stott et al.(2011)]{Stott2011} { Stott, J.~P., Collins, C.~A., Burke, C. et al., 2011, MNRAS, 414, 445}
\bibitem[Stott et al.(2012)]{Stott2012} { Stott, J.~P., Hickox, R.~C., Edge, A.~C. et al., 2012, MNRAS, 422, 2213}
\bibitem[Tacchella et al.(2016a)]{Tacchella2016a} Tacchella, S., Dekel, A., Carollo, C.~M. et al. 2016a, MNRAS, 457, 2790
\bibitem[Tacchella et al.(2016b)]{Tacchella2016b} Tacchella, S.,  Dekel, A., Carollo, C.~M. et al. 2016b, MNRAS, MNRAS, 458, 242
\bibitem[Tacconi et al.(2010)]{Tacconi2010} Tacconi, L. J., Genzel, R., Neri, R., et al. 2010, Nature, 463, 781
\bibitem[Tacconi et al.(2018)]{Tacconi2018} Tacconi, L.~J., Genzel, R., Saintonge, A., et al. 2018, ApJ, 853, 179
\bibitem[Tadaki et al.(2012)]{Tadaki2012} { Tadaki, K., Kodama, T., Ota, K. et al. 2012, MNRAS, 423, 2617}
\bibitem[Tadaki et al.(2014)]{Tadaki2014} { Tadaki, K., Kodama, T., Tamura, Y. et al. 2014, ApJ, 788, 23}
\bibitem[Tadaki et al.(2019)]{Tadaki2019} { Tadaki, K., Kodama, T., Hayashi, M. et al. 2019, PASJ, 71, 40}
\bibitem[Tan et al.(2014)]{Tan2014} { Tan, Q., Daddi, E., Magdis, G. et al. 2014,  A\&A, 569, 98}
\bibitem[Tanaka et al.(2013)]{Tanaka2013} { Tanaka, M., Toft, S., Marchesini, D. et al. 2013, ApJ, 772, 113}
\bibitem[Tran et al.(2010)]{Tran2010} { Tran, K.-V.~H., Papovich, C., Saintonge, A. et al. 2010,  ApJ, 719, 126}
\bibitem[Tremblay et al.(2016)]{Tremblay2016} { Tremblay, G.~R., Oonk, J.~B.~R., Combes, F. et al. 2016, Nature, 534, 218}
\bibitem[Trudeau et al.(2019)]{Trudeau2019} { Trudeau, A., Webb, T., Hlavacek-Larrondo, J. et al. 2019, MNRAS, 487, 1210}
\bibitem[Umehata et al.(2015)]{Umehata2015} { Umehata, H., Tamura, Y., Kohno, K. et al. 2015, ApJ, 815, 8}
\bibitem[van der Wel et al.(2012)]{vanderWel2012} { van der Wel, A., Bell, E.~F., H\"{a}ussler, B. et al. 2012, ApJS, 203, 24}
\bibitem[van der Wel et al.(2014)]{vanderWel2014}  van der Wel, A., Franx, M., van Dokkum, P. G., et al. 2014, ApJ, 788, 28
\bibitem[von der Linden et al.(2007)]{von_der_Linden2007} { von der Linden, A., Best, P.~N., Kauffmann, G. et al. 2007, MNRAS, 379, 867}
\bibitem[Wagg et al.(2012)]{Wagg2012} Wagg, J., Pope, A., Alberts, S. et al. 2012, ApJ, 752, 91 
\bibitem[Walter et al.(2012)]{Walter2012} { Walter, F., Decarli, R., Carilli, C. et al. 2012,  Nature, 486, 233}
\bibitem[Wang et al.(2016)]{Wang2016} { Wang, T., Elbaz, D., Daddi E. et al. 2016, ApJ, 828, 56}
\bibitem[Wang et al.(2018)]{Wang2018} { Wang, T., Elbaz, D., Daddi, E. et al. 2018, ApJ, 867, 29}
\bibitem[Webb et al.(2013)]{Webb2013} { Webb, T.~M.~A., O'Donnell, D., Yee, H.~K.~C. et al. 2013, AJ, 146, 84}
\bibitem[Webb et al.(2015a)]{Webb2015} { Webb, T.~M.~A., Muzzin, A., Noble, A. et al. 2015a, ApJ, 814, 96}
\bibitem[Webb et al.(2015b)]{Webb2015b} { Webb, T.~M.~A., Noble, A., DeGroot, A. et al. 2015b, ApJ, 809, 173}
\bibitem[Webb et al.(2017)]{Webb2017} Webb, T.~M.~A., Lowenthal, J., Yun, M. et al. 2017, ApJ, 844, 17
\bibitem[Whitaker et al.(2017)]{Whitaker2017} Whitaker, K. E., Bezanson, R., van Dokkum, P. G., et al. 2017, ApJ, 838, 19
\bibitem[White(1976)]{White1976} White, S.~D.~M., 1976, MNRAS, 177, 717
\bibitem[Wilson et al.(2009)]{Wilson2009} Wilson, G., Muzzin, A., \& Yee, H.~K.~C. 2009, ApJ, 698, 1943
\bibitem[Wright et al.(2017)]{Wright2017} { Wright, A.~H., Robotham, A.~S.~G., Driver, S.~P. et al. 2017, MNRAS, 470, 283 }
\bibitem[Young et al.(1995)]{Young1995} { Young, J.~S., Xie, S., Tacconi, L. et al. 1995, ApJS, 98, 219}
\bibitem[Yu et al.(2018)]{Yu2018} { Yu, H., Tozzi, P., van Weeren, R. et al. 2018, ApJ, 853, 100}
\bibitem[Zeimann et al. (2013)]{Zeimann2013} { Zeimann, G.~R., Stanford, S.~A., Brodwin, M. et al. 2013, ApJ, 779, 137}
\bibitem[Zhang et al.(2016)]{Zhang2016} Zhang, Y., Miller, C., McKay, T., et al. 2016, ApJ, 816, 98
\bibitem[Zirbel(1996)]{Zirbel1996} Zirbel, E.~L. 1996, ApJ, 473, 713
\bibitem[Zolotov et al.(2015)]{Zolotov2015} Zolotov, A., Dekel, A., Mandelker, N. et al. 2015, MNRAS, 450, 2327
\end{thebibliography}
\end{document}